\documentclass[prd,aps,12pt,showpacs,showkeys,nofootinbib]{revtex4-1}
\usepackage{graphicx}
\DeclareTextCommandDefault{\textcopyright}{\textcircled{c}}

\begin{document}

\title{Model of a multiverse providing the dark energy\\
  of our universe\footnote{\mbox{Preprint of an article published in
      Int.~J.~Mod.~Phys.~A, Vol. 32, No. 25 (2017) 1750149 (29 pages),\\}
    https://doi.org/10.1142/S0217751X17501494~\textcircled{c}~copyright~World~Scientific~Publishing~Company,\\
    http://www.worldscientific.com/worldscinet/ijmpa.\vspace{0.5\baselineskip}}}

\author{E.~Rebhan\footnote{Eckhard.Rebhan@uni-duesseldorf.de}}
\address{Institut f{\"u}r Theoretische Physik,\\
   Heinrich--Heine--Universit{\"a}t,\\
   D--40225 D{\"u}sseldorf, Germany}

\begin{abstract}
  It is shown that the dark energy presently observed in our universe
  can be regarded as the energy of a scalar field driving an
  inflation-like expansion of a multiverse with ours being a
  subuniverse among other parallel universes. A simple model of this
  multiverse is elaborated: Assuming closed space geometry, the origin
  of the multiverse can be explained by quantum tunneling from
  nothing; subuniverses are supposed to emerge from local fluctuations
  of separate inflation fields. The standard concept of tunneling from
  nothing is extended to the effect that in addition to an
  inflationary scalar field, matter is also generated, and that the
  tunneling leads to an (unstable) equilibrium state.  The
  cosmological principle is assumed to pertain from the origin of the
  multiverse until the first subuniverses emerge. With increasing age
  of the multiverse, its spatial curvature decays exponentially so fast
  that, due to sharing the same space, the flatness problem of our
  universe resolves by itself. The dark energy density imprinted by
  the multiverse on our universe is time-dependent, but such that the
  ratio $w{=}\varrho/(c^2p)$ of its mass density and pressure (times
  $c^2$) is time-independent and assumes a value $-1{+}\epsilon$ with
  arbitrary $\epsilon{>}0$. $\epsilon$ can be chosen so small, that
  the dark energy model of this paper can be fitted to the current
  observational data as well as the cosmological constant model.
\end{abstract}


\maketitle

\section{Introduction}
The concept of dark energy (DE) was introduced in order to fill a gap
in the present energy content of our universe and to explain the
presently observed acceleration of its
expansion~\cite{Perlmutter}. Best agreement between observations and
the predictions of the standard model of cosmology is obtained by
assuming that DE contributes 68,3 \% of the total energy in the
observable universe. The oldest and best known models for its
theoretical treatment are its representation by a cosmological
constant or by a dynamical scalar field, sometimes called quintessence
(see e.g. Ref.~\cite{Weinberg} and references therein). Numerous
further models were proposed, and still many more papers on DE have
been published. According to Ref.~\cite{Wang}, the proposed models can
be categorized into eight different groups, applications of the
holographic principle and back-reaction of gravity being more recent
ones. In spite of all these efforts, the true nature of DE is not yet
unveiled.

A scalar field used for describing cosmic inflation, called inflaton,
is of similar kind (obeys the same equations) as the scalar fields
used for modeling DE, save that it is much stronger. On the one hand,
inflation has, for many reasons, become an indispensable supplement to
the model for any kind of expanding universe. On the other hand,
important as it may be for our universe, DE seems to be no necessary
ingredient of a universe in general. In other words, DE is present in
our universe, but it is not clear why. This paper attempts to give an
answer to this question and therefore it cannot be classified among
the above mentioned categories in an important aspect. Technically, it
describes the DE observed in our universe by a scalar field. The
energy of the latter is shown to be identifiable with that of a scalar
field driving the inflation-like expansion of an all-embracing and
much older multiverse with ours being a subuniverse among others. Thus,
the answer to the above question is that our DE constitutes the
fingerprint of a superordinated multiverse generated by this energy.

The notion of a cosmic multiverse came up in the context of eternal
inflation, introduced by Steinhardt in Ref.~\cite{Gibbons} and by
Vilenkin~\cite{Vilenkin_1}. Thereby, the multiverse emerges as a
byproduct of the evolution of our universe from an inflaton field by
inflationary expansion and subsequent decay of the latter; actually, the
decay does not occur simultaneously everywhere, but at different times
in different places. The continuously growing number of subuniverses
thus produced is surrounded by rapidly growing regions of still
inflating space.

The approach of this paper is different in such a way that the
multiverse is considered as an independent entity, sustained by a
scalar inflation field $\Phi$ of its own which gives rise to a
permanent inflation-like expansion driven by an appropriately chosen
potential $V(\Phi)$. Within the multiverse, subuniverses with ours
among them are supposed to emerge from local fluctuations of separate
inflation fields, and for simplicity, it is assumed that as in hybrid
inflation (see Ref.~\cite{Linde1}), they do not inflate
eternally. (For better distinction from the latter and due to its --
further established -- identification with the DE of our universe,
$\Phi$ is called DE field in this paper.)  Furthermore, we assume that
the multiverse lives in curved closed space. Thus, similar to the case
of de Sitter space elaborated by Vilenkin~\cite{Vilenkin1,Vilenkin2},
its creation out of nothing becomes possible. An additional motivation
for assuming spatial closeness, based on a not quite obvious but
profound difference between closed and open space, is presented in
Appendix~\ref{app:1}.\footnote{A similar model of an expanding
  multiverse with flat spatial geometry would be possible, but appear
  less meaningful, because for the reason of causal connectivity, the
  multiverse could only cover a finite region of the infinitely
  extended space-time. A creation out of nothing would be impossible,
  and the argument of Appendix~\ref{app:1} would not apply.}%
~For the sake of simplicity we assume that the closed multiverses
considered satisfy the cosmological principle from their origin until
the first subuniverses emerge.

A closed space-time with Friedmann Robertson Walker (FRW) metric,
generated by a DE field $\Phi$ alone represents a homogeneous and
isotropic entity that in the following is occasionally referred to as
$\Phi$-multiverse.  In Section~\ref{sec:accel-exp}, it is shown that
\mbox{$\Phi$-multi}\-verses can provide the background of
inhomogeneous multiverses containing a multitude of subuniverses, what
means, that $\Phi$ keeps to be the only cosmic substrate and remains
unaffected (all internal symmetries being maintained) within all space
between the gradually emerging subuniverses. Due to this comprehensive
role of the $\Phi$-multiverses, an important task of the present paper
consists in deriving suitable solutions for them (Section~\ref{sec:Evo
  without}).

The property of never-ending inflation-like expansion of the
multiverse requested above can be achieved by assuming that the
potential $V(\Phi)$ decreases monotonically with increasing $\Phi$ for
all values of $\Phi$. Since then, there is no minimum of $V(\Phi)$
around which the field $\Phi$ could oscillate, a decay of the field
$\Phi$ by phase transition, triggered by field oscillations (see
e.g. p.~244 of Ref.~\cite{Mukhanov}), is avoided.

Simple as it may look at first glance, the multiverse concept outlined
above has many implications and requires a multitude of calculations
for establishing a consistent model. The following requirements must
be satisfied (the sections where care is taken of them being noted in
brackets):

\begin{enumerate}
\item\label{i2} $\Phi$ fulfills the general relativistic equations for
  a scalar quantum field, driven by a potential $V(\Phi)$ which
  prevents oscillations of $\Phi(T)$ and causes a continuously
  accelerated inflation-like expansion (Section~\ref{sec:Evo without}).
\item\label{i3} The DE presently observed in our universe can be
  attributed to the field $\Phi$. For that, the mass density
  $\varrho_{\Phi}$ of $\Phi$ must not only have the proper present
  value (in Section~\ref{sec:Relations}, it is derived how it
  transfers to our universe), but must also drive the accelerated
  expansion presently observed in our universe
  (Section~\ref{sec:infl-sub}). Furthermore it may not be
  space-dependent there (Section~\ref{sec:Relations}).
\item\label{i4} The initial value of $\varrho_\Phi$ may not exceed
  the Planck density $\varrho_P$ (Section~\ref{sec:gen-theo}).
\item\label{i5} Our universe fits into the multiverse, not only
  spatially but also time-wise (Section~\ref{sec:eternal}).
\item\label{i6} The present curvature of the multiverse is so small
  that it lies well below the present limits of measurability
  (Section~\ref{sec:eternal}).  
\item\label{i7} The dynamics of the expansion can be arranged in such
  a way that the initial state can be explained to come about by
  quantum tunneling from nothing  (Section~\ref{sec:quant-tunn}).
\item\label{i8} The properties of our universe and of the DE following
  from our model agree with those predicted by the standard model of
  cosmology for a practically uncurved space and are fitting the
  current observational data (Section~\ref{sec:infl-sub}
  and~\ref{sec:obs-const}).
\end{enumerate}

Since the number of model parameters is quite small for a fitting, it
is by no means evident that all requirements can be satisfied.

\section{Evolution of pure $\Phi$-multiverses}
\label{sec:Evo without}

In this section, we determine the evolution of $\Phi$-multiverses
ignoring the presence of subuniverses. According to
Section~\ref{sec:account_sub}, in a multiverse with subuniverses this
yields the correct result for all regions outside the latter ones. All
calculations are carried out in FRW coordinates. The basic equations
to be satisfied in a closed multiverse with positive spatial curvature
are (see, e.g., Ref.~\cite{Weinberg1} or p. 550 in
Ref.~\cite{Rebhan1})%
\footnote{In order to make numerical evaluations more transparent, in
  this paper MSI units are used.}
\begin{eqnarray}
  H^2 =  \frac{\dot{A}^2(T)}{A^2(T)} 
  = \frac{8\pi G}{3}\varrho -\frac{c^2}{A^2}
  && \qquad\mbox{with}\qquad 
  \varrho = \varrho_{m} + \varrho_{\Phi}\,,
  \label{eq:0.1}\\ 
  \varrho_{\Phi} =  \frac{\hbar^2 \dot{\Phi}^2(T)}{2\mu c^4}+ \frac{V(\Phi)}{c^2}\,,
                  \quad  && \quad
      p_{\Phi} = \frac{\hbar^2 \dot{\Phi}^2(T)}{2\mu c^2} - V(\Phi)\,,
  \label{eq:0.2}\\   
  \ddot{\Phi}(T) + 3H\dot{\Phi}(T) + \frac{\mu c^2}{\hbar^2}\,V'(\Phi) &=& 0\,.
  \label{eq:0.3}              
\end{eqnarray}
For better discriminability, the cosmic scale factor and the time are
denoted by $A$ and $T$ in a multiverse and by $a$ and $t$ in
subuniverses. $\varrho_{m}$ and $\varrho_{\Phi}$ are the mass
densities of matter and the DE field $\Phi$, respectively, and $\mu$ is
the mass parameter of the field~$\Phi$.

For the evaluation of Eqs.~(\ref{eq:0.1})-(\ref{eq:0.3}), it turns out
useful to introduce relative quantities, and to do this separately for
the very early and the later evolution of the multiverse; furthermore,
we admit the possibility that initially besides the DE field, matter
is also present. Since the present value of $\varrho_\Phi$ plays a
decisive role, we begin with the later evolution.

\vspace*{-0.5\baselineskip}
\subsection{Later evolution}
\label{sec:late-ev}

\subsubsection{General theory}
\label{sec:gen-theo}

The relative quantities used for the later evolution are
\begin{equation}
  \label{eq:1.1}
  x = A/A_0\,,\qquad\qquad   \tau = T/t_{H0}\,,
\end{equation}
the index zero referring to the present values, i.e. $A_0{=}A(T_0)$
etc. and $x{=}1$ for $T{=}T_0$.
\begin{equation}
  \label{eq:1.2}
  t_{H0} = \sqrt{\frac{3}{8\pi G \varrho_{c0}}} = \frac{1}{H_0} 
  = 14.0\cdot 10^{9}\,\mbox{a} 
  = 4.41\cdot 10^{17}\,\mbox{s}
\end{equation}
 and
\vspace*{-0.5\baselineskip}
\begin{equation}
  \label{eq:1.3}
  \varrho_{c0} = \frac{3H_0^2}{8\pi G} 
  = 9.20\cdot 10^{-27}\,\mbox{kg\,m}^{-3}
\end{equation}
are the present Hubble time and the critical density of our universe
respectively for the Hubble parameter%
\footnote{Latest measurements yielded $H_0{=}67.6^{+0.7}_{-0.6}$
  (SDSS-III Baryon Oscillation Spectroscopic Survey data from
  07.13.2016) and $H_0{=}71.9^{+2.4}_{-3.0}$ (Hubble Space Telescope
  data from 11.22.2016).}%
~$H_0{=}70\,\mbox{km\,s}^{-1}/\mbox{Mpc}$. Considering the densities
as functions of $x$ we set
\begin{equation}
  \label{eq:1.4}
  \varrho = \varrho_{\Phi 0}\,g(x)
  \qquad\mbox{with}\qquad
  g(x) = \frac{\varrho_m(A_0x)+\varrho_\Phi(A_0x)}{\varrho_{\Phi0}}\,.
\end{equation}
According to the later Eq.~(\ref{eq:2.5}), $\varrho_{\Phi 0}$ can
be identified directly with the mass density $0.683\,\varrho_{c0}$ of
DE presently observed in our universe, i.e.
\begin{equation}
  \label{eq:1.5}
  \varrho_{\Phi 0} = \varrho_{\Phi}(A_0)= 0.683\,\varrho_{c0}
  = 6.28\cdot 10^{-27} \,\mbox{kg\,m}^{-3}\,.
\end{equation}
Inserting Eqs.~(\ref{eq:1.1})-(\ref{eq:1.4}) in Eq.~(\ref{eq:0.1})
yields
\begin{equation}
  \label{eq:1.6}
  \dot{x}^2(\tau) =  \frac{\varrho_{\Phi 0}}{\varrho_{c0}}\,
  \left(x^2g(x) - \frac{3c^2}{8\pi G\varrho_{\Phi 0}\,A_0^2}\right)\,.
\end{equation}
Since our universe must fit into the multiverse, $A_0\gg R$ must hold
where 
\begin{equation}
  \label{eq:1.7}
  R = a(t_0)\,r_{bo} = 23.5\cdot 10^{9}\,\mbox{ly} = 2.22\cdot 10^{26}\,\mbox{m}
\end{equation}
is the present metric radius of the boundary of our observable
universe (at radial coordinate $r_{bo}$ and for scale factor $a(t_0)$
reached at present time $t_0$). Accordingly, the second term in the
brackets is very small, and it turns out, that in all interesting cases,
the bracket has a zero at some $x{=}x_*{<}1$. This zero is admissible,
if the physical solution is prevented from running into it. The latter
can be achieved by requiring $x_*\leq x_i$, where $x_i$ is the initial
value from which the expansion of the multiverse starts. We decide
for the choice $x_*{=}x_i$ or 
\begin{equation}
  \label{eq:1.8}
  x_i^2\,g(x_i) = \frac{3c^2}{8\pi G\varrho_{\Phi 0}\,A_0^2}\,,
\end{equation}
which causes the multiverse to start with zero expansion velocity,
\begin{equation}
  \label{eq:1.9}
  \dot{x}(\tau) = 0 \qquad\mbox{for}\qquad
  x = x_i\,,
\end{equation}
and enables its creation out of nothing by quantum tunneling (see
Subsection~\ref{sec:quant-tunn}). With this and Eq.~(\ref{eq:1.5}),
Eq.~(\ref{eq:1.6}) finally yields
\begin{equation}
  \label{eq:1.10}
  \dot{x}(\tau) =0.826\,\sqrt{x^2\,g(x)-x_i^2\,g(x_i)}
\end{equation}
and
\vspace*{-\baselineskip}
\begin{equation}
  \label{eq:1.11}
  \tau(x) = 1.21\,\int_{x_i}^x\frac{d\xi}{\sqrt{\xi^2\,g(\xi)-x_i^2\,g(x_i)}}
  \qquad\mbox{for}\qquad
  \tau(x_i) = 0\,.
\end{equation}
We assume that the expansion of the multiverse starts from the scale
factor $A_i{=}l_P$, where
\begin{equation}
  \label{eq:1.12}
  l_{P} = \sqrt{\frac{\hbar G}{c^3}} = 1.616\cdot 10^{-35}\,\mbox{m}
\end{equation}
is the Plank length, whence we have
\begin{equation}
  \label{eq:1.13}
  x_i = \frac{l_P}{A_0} = \frac{l_P}{R\,\zeta}
  \qquad\mbox{with}\qquad
  \zeta:=\frac{A_0}{R}\,.
\end{equation}
(In Section~\ref{sec:obs-const} consequences of the more general
condition $A_i{=}\lambda l_P$ with $\lambda{\geq}1$ are discussed.)
Inserting this in Eq.~(\ref{eq:1.8}) yields
\begin{equation}
  \label{eq:1.14}
  g_i = g(x_i) = \frac{3c^2}{8\pi G\varrho_{\Phi 0}\,l_P^2} =
  9.80\cdot 10^{121}
  \qquad\mbox{and}\qquad
  \zeta =\frac{l_P}{R\,x(g_i)} = \frac{7.28\cdot 10^{-62}}{x_i}\,.
\end{equation}
(For evaluation $c{=}2.998\cdot10^8\,\mbox{ms}^{-1}$,
$G{=}6.673\cdot 10^{-11}\,\mbox{m}^3\mbox{kg}^{-1}\mbox{s}^{-2}$ and
Eqs.~(\ref{eq:1.5}), (\ref{eq:1.7}) and (\ref{eq:1.12}) were used;
$x(g)$ is the inverse of the function $g(x)$.) Imposing on $g(x)$ the
condition
\begin{equation}
  \label{eq:1.15}
  d[x^2 g(x)]/dx > 0
  \qquad\mbox{for~all}\quad x \,, 
\end{equation}
according to Eq.~(\ref{eq:1.10}) we achieve a continuously accelerated
expansion of the multiverse that with growing $x$ approaches an
inflation-like state.

For checking the fulfillment of requirement~\ref{i4} of the
Introduction, the initial density $\varrho_i$ must be calculated using
the initial condition~(\ref{eq:1.9}) or $\dot{A}(T){=}0$ for
$A{=}A_i{=}l_P$. For this, we obtain from Eqs.~(\ref{eq:0.1}),
(\ref{eq:1.4}a) and~(\ref{eq:1.12})
\begin{equation}
  \label{eq:1.16}
  \varrho_i =  \varrho_{mi} +  \varrho_{\Phi i} = \frac{3c^2}{8\pi G\,l_P^2}
  = \frac{3}{8\pi}\,\varrho_P =  \varrho_{\Phi 0}\,g_i\,,
\end{equation}
where at last the definition%
\vspace*{-0.5\baselineskip}
\begin{equation}
  \label{eq:1.17}
  \varrho_{P} =\frac{c^5}{\hbar\,G^2} = 5.157\cdot 10^{96}\,\mbox{kg~m}^{-3}
\end{equation}
of the Planck density $\varrho_{P}$ was used. $\varrho_i$ is fixed to
a value slightly below the Planck density as demanded. This result is
completely independent of the composition and later behavior of the
density~$\varrho$ and is solely due to the vanishing of the initial
expansion velocity.

We now turn to solving Eqs.~(\ref{eq:0.2})-(\ref{eq:0.3}). Inserting
$\varrho_\Phi{=}\varrho_\Phi(A)$ in Eq.~(\ref{eq:0.2}a)%
\footnote{Eq.~(\ref{eq:0.2}a) is supposed to denote the first of the
  Eqs.~(\ref{eq:0.2}), Eq.~(\ref{eq:0.2}b) the second etc..}%
~and deriving it with respect to $T$ yields%
\vspace*{-0.5\baselineskip}
\begin{equation}
  \label{eq:1.18}
  \;\dot{\Phi}(T)\,\left(\frac{\hbar^2 \ddot{\Phi}(T)}{\mu c^4} 
  + \frac{V'(\Phi)}{c^2}\right) = \varrho'_\Phi(A)\,\dot{A}(T)\,.
\end{equation}
Inserting in this $\ddot{\Phi}(T)$ from Eq.~(\ref{eq:0.3}) and
$H{=}\dot{A}(T)/A$, after some rearrangement we obtain
\begin{equation}
  \label{eq:1.19}
  \dot{\Phi}(T) = \pm\frac{c^2}{\hbar}\,
  \sqrt{-\frac{\mu A\varrho'_\Phi(A)}{3}}\,.
\end{equation}
Note that this equation holds independent of whether or not
$\varrho_m{\equiv}0$. From it follows the conditions
$\varrho'_\Phi(A){\leq}0$. Since according to condition (\ref{i2}) of
the Introduction $\Phi(T) $ may not oscillate, the possibility
$\dot{\Phi}(T){=}0$ must be excluded whence
\begin{equation}
  \label{eq:1.20}
  \varrho'_\Phi(A) < 0
  \qquad\mbox{and}\qquad
  \dot{\Phi}(T) = \frac{c^2}{\hbar}\,
  \sqrt{-\frac{\mu A\varrho'_\Phi(A)}{3}}\,.
\end{equation}
(Due to $\dot{\Phi}(T){\neq}0$, the minus branch of $\dot{\Phi}(T)$
displayed in Eq.~(\ref{eq:1.19}) can be excluded.) Introducing $x$
and $\tau$ from Eqs.~(\ref{eq:1.1}) in Eq.~(\ref{eq:1.20}) yields
\begin{equation}  
  \label{eq:1.21}
  \dot{\Phi}(\tau) = \delta\,\sqrt{-x\,f'(x)}
  \qquad\mbox{where}\qquad
  f(x) = \frac{\varrho_\Phi(A_0x)}{\varrho_{\Phi 0}}\,, \quad 
  \delta = \frac{c^2t_{H0}\sqrt{\mu\varrho_{\Phi 0}}}{\sqrt{3}\,\hbar}\,.
\end{equation}
From Eqs.~(\ref{eq:1.10}) and (\ref{eq:1.21}a), we get
\begin{equation}
  \label{eq:1.22}
  \frac{d\Phi}{dx} =  \frac{\dot{\Phi}(\tau)}{\dot{x}(\tau)} 
  =  \frac{1.21\,\delta\,\sqrt{-x\,f'(x)}}{\sqrt{x^2\,g(x){-}x_i^2\,g(x_i)}}
  \quad\mbox{or}\quad  
  \Phi(x) = 1.21\,\delta\!\int_{x_i}^x\!\frac{\sqrt{-\xi\,f'(\xi)}\,d\xi}
  {\sqrt{\xi^2\,g(\xi){-}x_i^2\,g(x_i)}}
\end{equation}
for the initial condition $\Phi(x_i){=}0$. Inserting
Eqs.~(\ref{eq:1.19}) and~(\ref{eq:1.21}b) in Eq.~(\ref{eq:0.2}a)
yields
\begin{equation}
  \label{eq:1.23}  
  \frac{V(\Phi)}{\varrho_{\Phi 0}\,c^2} 
  = \left[f(x)+\frac{x\,f'(x)}{6}\right]_{x=x(\Phi)}\,,
\end{equation}
where $x(\Phi)$ is the inverse function of $\Phi(x)$. Although
Eq.~(\ref{eq:0.3}) was used in deriving Eq.~(\ref{eq:1.23}), we must
still make sure that it is actually satisfied. According to
Eqs.~(\ref{eq:1.18}) and (\ref{eq:0.1}) we have
\begin{displaymath}
  \ddot{\Phi}(T)+\frac{\mu c^2}{\hbar^2} V'(\Phi) 
  = \frac{\mu c^4A\varrho'_\Phi(A)}{\hbar^2}\,\frac{H}{\dot{\Phi}(T)}
  = -3H\dot{\Phi}(T)\,,
\end{displaymath}
where at last the square of Eq.~(\ref{eq:1.19}) was used. This
confirms the fulfillment of Eq.~(\ref{eq:0.3}) and reveals together
with the result~(\ref{eq:1.23}) that the field $\Phi$ has the
structure required by the usual theory of scalar fields
(requirement~\ref{i2} of the Introduction).

\subsubsection{Specialization to  
  $\varrho_{\Phi}\sim x^{\gamma-2}$}
\label{sec:eternal}

Due to our assumption that the field $\Phi$ does not decay into regular
matter by a phase transition, we must only account for initially
present mass of density $\varrho_m$ obeying the equation of state%
\vspace*{-0.5\baselineskip}
\begin{equation}
  \label{eq:1.24}
  \varrho_m = \frac{\varrho_{mi}A_i^n}{A^n} = \frac{\varrho_{mi}x_i^n}{x^n}\,,
\end{equation}
where $n{=}3$ for cold and $n{=}4$ for hot matter. Concerning an
appropriate ansatz for $\varrho_{\Phi}$, we have to observe the
inequality~(\ref{eq:1.20}a) or $g'(x){<}\,0$, and the
inequality~(\ref{eq:1.15}), according to which $g(x)$ must decrease
more slowly than $\hat{g}(x){=}C/x^2$ with increasing $x$ since
$d[x^2\hat{g}(x)]/dx{=}0$. Due to this and the validity of
Eq.~(\ref{eq:1.26}a) for almost all $x$, we employ the ansatz
\begin{equation}
  \label{eq:1.25}
  \varrho_{\Phi} = \varrho_{\Phi 0}\,x^{\gamma-2}
  \qquad\mbox{or}\qquad
  f(x) = x^{\gamma-2} 
  \qquad\mbox{with}\qquad
  0 < \gamma < 2 \,.
\end{equation}
Since the spatial curvature of the $\Phi$-multiverse is
$K{=}1/A^2{\sim}\,x^{-2}$, we have
$\varrho_{\Phi}{\sim}K^{1-\gamma/2}$. This means, that the mass
density $\varrho_{\Phi}$ of the field $\Phi$ is coupled to the
curvature $K$ in a monotonic way such, that it decreases and
approaches to zero together with the latter. This property goes well
with the considerations \mbox{of~\ref{app:1}}, according to which in a
closed multiverse of positive curvature an intrinsic expansion of
space should exist, not present in an open multiverse.

Restricting the initial value of $\varrho_m$ to the value specified in
Eq.~(\ref{eq:1.43a}b) and denoting the corresponding initial value of
$x$ by $x_{im}$, we get from Eqs.~(\ref{eq:1.24}) and~(\ref{eq:1.25}a)
\begin{displaymath}
  \frac{\varrho_m}{\varrho_\Phi} =
  \frac{\gamma}{n{-}2}\left(\frac{x_{im}}{x}\right)^{n+\gamma-2}
    \ll 1
  \qquad\mbox{for}\qquad
  x \gg x_{im}\left(\frac{\gamma}{n{-}2}\right)^{\frac{1}{n+\gamma-2}}
    =  x_{im}\cdot\mathcal{O}(1)\,. 
\end{displaymath}
For example, for $\gamma{=}1.95$ and $n{=}3$ from
Eqs.~(\ref{eq:1.29}a) and (\ref{eq:1.30}a), we get
$x_{im}{=}\mathcal{O}(1){\cdot}10^{-2431}$. It can be concluded from
this, that for almost the entire later evolution $\varrho_m$ can be
neglected, whence according to Eqs.~(\ref{eq:1.4}a)
and~(\ref{eq:1.21}b) we get
\begin{equation} 
  \label{eq:1.26}
  g(x) = f(x) \qquad\mbox{and}\qquad
  f(1) = 1 \,;
\end{equation}
in other words the later evolution is practically the same for the
cases with and without primordial matter. With Eq.~(\ref{eq:1.25}) and
this, Eqs.~(\ref{eq:1.10})-(\ref{eq:1.11}) become
\begin{equation}
  \label{eq:1.27}
  \dot{x}(\tau) = 0.826\,\sqrt{x^\gamma-x_i^\gamma}
\end{equation}
and
\begin{equation}
  \label{eq:1.28}  
   \tau(x) = 1.21\!\int_{x_i}^x\!\frac{d\xi}{\sqrt{\xi^{\gamma}-x_i^\gamma}}
   = -\frac{1.21\,x \sqrt{x^\gamma{-}x_i^\gamma}}{x_i^\gamma}\,
   \,_2F_1\left(1,\frac{2{+}\gamma}{2\,\gamma};
    \frac{1{+}\gamma}{\gamma};\frac{x^\gamma}{x_i^\gamma}\right)\,,
\end{equation}
where $\,_2F_1(a,b;c;z)$ is the hypergeometric function. The
quantities $x_i$ and $\zeta$, for $\varrho_m{\neq}0$ and restriction
to the special case of Eq.~(\ref{eq:1.43a}b) denoted by $x_{im}$ and
$ \zeta_m$, follow from Eqs.~(\ref{eq:1.14}) with the use of
Eqs.~(\ref{eq:1.4}) and (\ref{eq:1.25}a) and are
\begin{eqnarray}  
   \hspace*{-8mm}   x_i =  x(g_i) =  g_i^{-\frac{1}{2-\gamma}}\!,\qquad && \qquad
              \,\zeta =  \mbox{e}^{-140.775+\frac{280.895}{2-\gamma}}
              \qquad\;\mbox{for}\;\;\varrho_{mi}=0\,,\label{eq:1.29}\\
  \hspace*{-6mm}  x_{im} = x_i\left(\frac{n{+}\gamma{-}2}{n{-}2}\right)^{\!\frac{1}{2-\gamma}}\!\!,
  \hspace*{2mm}  &&  \qquad
      \zeta_m = \zeta\left(\frac{n{-}2}{n{+}\gamma{-}2}\right)^{\!\frac{1}{2-\gamma}}\;
                \,\;\mbox{for}\;\; \varrho_{mi}=\frac{\gamma\,\varrho_{\Phi i}}{(n{-}2)}
                \quad \label{eq:1.30} 
\end{eqnarray}
with $g_i$ given by Eq.~(\ref{eq:1.14}a). A very good approximation to
the result~(\ref{eq:1.28}) is obtained by expanding the integral in
Eq.~(\ref{eq:1.28}) with respect to $u^{-\gamma/2}$ where
$u{=}\xi/x_i$:
\begin{eqnarray}
  \frac{\tau(x)}{1.21}&=&x_i^{1-\frac{\gamma}{2}}\!\int_1^{\frac{x}{x_i}}\!
                          \frac{u^{-\frac{\gamma}{2}}\,du}{\sqrt{1{-}u^{-\gamma}}}
                          = x_i^{1-\frac{\gamma}{2}} \!\int_1^{\frac{x}{x_i}}\!
                          \left(u^{-\frac{\gamma}{2}}+\frac{u^{-\frac{3\gamma}{2}}}{2}
                          +\frac{3\,u^{-\frac{5\gamma}{2}}}{8}+\dots\right)du
                          \nonumber\\
                        &=&\frac{x^{1-\frac{\gamma}{2}}}{1{-}\gamma/2}
                        -Z(\gamma)\,x_i^{1-\frac{\gamma}{2}}
                            +\mathcal{O}(x_i^\gamma) \nonumber
\end{eqnarray}
with%
\vspace*{-\baselineskip}
\begin{displaymath}
  Z(\gamma)=\frac{2}{2{-}\gamma}+\frac{1}{2{-}3\gamma}+
  \frac{3}{4\,(2{-}5\gamma)} +\dots\,.
\end{displaymath}
Due to the extreme smallness of $x_i$, for
$x{\gg}[(1{-}\gamma/2)\,Z(\gamma)]^{1/(1-\gamma/2)}x_i{=}\mathcal{O}(1)\,x_i$
the result~(\ref{eq:1.28}) can be replaced
by
\begin{equation}
  \label{eq:1.31}
  \tau(x) = \tau_0\,x^{1-\gamma/2}
  \quad\mbox{with}\quad
  \tau_0=\frac{1.21}{1-\gamma/2}\,,
\end{equation}
the error being negligibly small. Equivalently we have
\begin{equation}
  \label{eq:1.32}
  x = (\tau/\tau_0)^{1/(1-\gamma/2)}\,.
\end{equation}
Inserting $\gamma{=}2{-}2.42/\tau_0$, obtained from
Eq.~(\ref{eq:1.31}b), in Eqs.~(\ref{eq:1.29}b) and~(\ref{eq:1.30}b)
yields
\begin{eqnarray}
  \zeta(\tau_0) &=& \mbox{e}^{-140.775+116.072\,\tau_0}  \label{eq:1.33}
  \\
  \zeta_m(\tau_0) &=& \zeta(\tau_0)\left(\frac{n}{n{-}2}
                      {-}\frac{2.42}{(n{-}2)\tau_0}\right)^{\!\!-0.413\,\tau_0}
                      \;
                      \renewcommand*\arraystretch{0.6} 
                      \begin{array}{c}
                        {\scriptstyle \tau_0\gg 1}\\
                        \longrightarrow
                      \end{array}
                      \quad
                      \zeta(\tau_0)\,\mbox{e}^{-0.413\,\tau_0\ln(\frac{n}{n{-}2})}.
     \label{eq:1.34}
\end{eqnarray}
From Eq.~(\ref{eq:1.33}) follows the condition
$\tau_0{>}140.775/116.072{=}1.21..$, since according to
condition~\ref{i5} of the Introduction and Eq.~(\ref{eq:1.13}b), we
must have $\zeta{>}1$ or $\zeta_m{>}1$ respectively. Due to the
exponential growth with $\tau_0$, already for $\tau_0{\approx}2.4$
both $\zeta$'s reach a $10^{60}$-fold enhancement above the demanded
minimum value $1$. This means that for almost all admissible ages of
the multiverse its present radius $A_0{=}R \zeta$ is so huge that the
present spatial curvature%
\vspace*{-0.5\baselineskip}
\begin{equation}
  \label{eq:1.34a}
  K_0 = \frac{1}{A_0^2} = \frac{1}{R^2\,\zeta^2}
  \vspace*{-0.5\baselineskip}
\end{equation}
is far below measurability.

\begin{table}
\caption{Parameters of the late multiverse as functions of $\gamma$ for $n{=}4$.}
{\begin{tabular}{@{}l|ccccccc@{}} \toprule
  $\gamma$ & $0.1$ & $1.0$ & $1.8$ & $1.9$ & $1.95$ & $1.99$ \\\colrule
    $\tau_0$ & $1.27$ & $2.42$ & $12.1$ & $24.2$ & $48.4$ & $242$ \\
    $\zeta$  & $1.2{\cdot} 10^3$ & $7.1{\cdot} 10^{60}$ & $ 6.6{\cdot}10^{548}$ &  
    $5.9{\cdot} 10^{1158}$ & $4.9{\cdot} 10^{2378}$ & $9.6{\cdot} 10^{12137}$ \\
    $\zeta_m$  & $1.1{\cdot} 10^3$ & $4.8{\cdot} 10^{60}$ & $ 2.7{\cdot}10^{547}$ &
    $7.5{\cdot} 10^{1155}$ & $6.0{\cdot} 10^{2372}$ & $9.7{\cdot} 10^{12107}$ \\
   $x_i$     & $6.2{\cdot}10^{-65}$ &$1.0{\cdot}10^{-122}$ &$1.1{\cdot}10^{-610}$&
            $1.2{\cdot}10^{-1220}$ &$1.5{\cdot}10^{-2440}$ &$7.5{\cdot}10^{-12200}$\\  
    $x_s$     & $0.213$ & $0.354$ & $0.430$ & $0.437$ & $0.441$ & $0.443$ \\  
    $\ddot{y}_0/\ddot{y}_\Lambda$ & $-0.021$ & $ 0.463$ & $0.893$ & $0.946$ &
    $ 0.973$ & $0.995$\\
    $\varrho_{\Phi h}/\varrho_{\Phi 0}\!\!$ & $2.64$ & $1.57$ & $1.09$ & $1.04$ &
    $1.02$ & $1.00$ \\\botrule
\end{tabular} \label{tab:1}}
\end{table}

In Appendic~\ref{app:3} it is shown that for $\gamma$ values close to
$2$, the condition~(\ref{app:3.1}) for slow roll is satisfied in the
whole range of validity of the later evolution.  Simultaneously, the
solution~(\ref{eq:1.36}) for $V$ approaches the slow roll
approximation~(\ref{app:3.2}),
$V(x){=}\varrho_{\Phi 0}\,c^2\,x^{\gamma-2}{=}\varrho_\Phi(x)\,c^2$.

In table~\ref{tab:1}, the present age $\tau_0{=}\tau(1)$ of the
multiverse is shown in multiples of $t_{H_0}$ for several values of
$\gamma$ together with the corresponding values of $\zeta$ and
$\zeta_m$ from Eqs.~(\ref{eq:1.29})-(\ref{eq:1.30}) and with further
values discussed in Section~\ref{sec:infl-sub}. It is seen that for
reasonable ages $\tau_0{\gg}1$ the parameter $\gamma$ must be close
to~$2$.

Inserting Eqs.~(\ref{eq:1.25}b) and (\ref{eq:1.26}a) in
Eq.~(\ref{eq:1.22}b) and substituting $u{=}\xi/x_i$ yields
\begin{displaymath}  
  \Phi = 1.21\,\delta\,\sqrt{2{-}\gamma}\int_{x_i}^x\!
  \sqrt{\frac{\xi^{\gamma-2}}{\xi^\gamma-x_i^\gamma}}\,d\xi
  =1.21\,\delta\,\sqrt{2{-}\gamma}\int_{1}^{\frac{x}{x_i}}\!
  \frac{du}{u\sqrt{1{-}u^{-\gamma}}}.
\end{displaymath}
Expanding $1/\sqrt{1{-}u^{-\gamma}}$ with respect to $u^{-\gamma}$ and
neglecting all terms ${\sim}\,(x_i/x)^{n\gamma}$, for $x{\gg}x_i$, we
obtain the approximate result
\begin{equation}  
  \label{eq:1.35}
  \Phi =1.21\,\delta\,\sqrt{2{-}\gamma}\left(\ln \frac{x}{x_i}+\frac{Z}{\gamma}\right) 
  \qquad\mbox{or}\qquad
  x = x_i\,\exp\left(\frac{0.826\,\Phi}{\delta\,\sqrt{2{-}\gamma}}
      -\frac{Z}{\gamma}\right)\,,
\end{equation}
where%
\vspace*{-\baselineskip}
\begin{displaymath}
  Z = \frac{1}{2}+\frac{3}{2\cdot 8}+\frac{5}{3\cdot 24}+\dots \approx 1.386\,.
\end{displaymath}
Inserting Eqs.~(\ref{eq:1.25}b) and~(\ref{eq:1.35}b) in
Eq.~(\ref{eq:1.23}) yields
\begin{equation}
  \label{eq:1.36}
  \frac{V(\Phi)}{\varrho_{\Phi 0}\,c^2} = \frac{4+\gamma}{6\,x^{2-\gamma}}
  = \frac{4+\gamma}{6\,x_i^{2-\gamma}}\,
  \exp\left( \frac{Z}{\gamma} -\frac{0.826\,\Phi}{\delta\,\sqrt{2{-}\gamma}}
     \right).
\end{equation}

\subsection{Very early evolution and creation of the multiverse}
  \label{sec:creation}

\subsubsection{Very early evolution}

For dealing with the early evolution of $\Phi$-multiverses, we adapt
our equations to the small scales involved. Instead of the relative
quantities defined in Eq.~(\ref{eq:1.1}), now we employ
\begin{equation}
  \label{eq:1.37}
  x = \frac{A}{l_P}\,,\qquad\quad
  \tau = \frac{T}{t_P}
  \qquad\mbox{with}\qquad
  t_P = \sqrt{\frac{\hbar G}{c^5}}=5.39\cdot 10^{-44}\,\mbox{s}\,,
\end{equation}
whence $x{=}1$ for $A{=}l_P$. Instead of the
Eqs.~(\ref{eq:1.24}) and (\ref{eq:1.25}) we use
\begin{equation}
  \label{eq:1.38}  
  \varrho_{\Phi} = \varrho_{\Phi i}\,x^{\gamma-2}\,,\qquad
  \varrho_{m} = \varrho_{m i}\,x^{-n}\,.
\end{equation}
With this and the identities $l_P/t_P{=}c$ and $G\,t_P^2\varrho_P{=}1$,
the latter following from the definitions~(\ref{eq:1.37}c) and
(\ref{eq:1.17}), Eq.~(\ref{eq:0.1}) becomes
\begin{equation}
  \label{eq:1.40}
  \dot{x}(\tau) = \sqrt{\frac{8\pi}{3\,\varrho_P}
  \left(\varrho_{mi}\,x^{2-n}+\varrho_{\Phi i}\,x^\gamma\right)-1}\,.
\end{equation}
In the case $\varrho_{mi}{=}0$, with Eq.~(\ref{eq:1.17}), we get
$\varrho_{\Phi i}{=}\varrho_i {=} 3\varrho_P/(8\pi)$ and
Eq.~(\ref{eq:1.40}) simplifies to
\begin{equation}
  \label{eq:1.42}
  \dot{x}(\tau) = \sqrt{x^\gamma -1}\,.
\end{equation}
Differentiating Eq.~(\ref{eq:1.40}) with respect to $\tau$ yields
\begin{equation}
  \label{eq:1.43}
  \ddot{x}(\tau) =  \frac{4\pi\left[(2{-}n)\,\varrho_{mi}\,x^{2-n}
      +\gamma\,\varrho_{\Phi i}\,x^\gamma\right]}{3\varrho_P x}\,,
\end{equation}
and from this it follows, that at $x{=}1$ we have
\begin{equation}
  \label{eq:1.43a}
  \ddot{x}(\tau) = 0   \qquad\mbox{for}\qquad
    \varrho_{mi} = \frac{\gamma\,\varrho_{\Phi i}}{n{-}2}\,.
\end{equation}
Inserting this in Eq.~(\ref{eq:1.16}) yields
\begin{equation} 
  \label{eq:1.44}
  \varrho_{mi} = \frac{3\varrho_P}{8\pi}\frac{\gamma}{(n{-}2{+}\gamma)}\,,
  \qquad
  \varrho_{\Phi i} = \frac{3\varrho_P}{8\pi}\frac{(n{-}2)}{(n{-}2{+}\gamma)}\,.
\end{equation}
For these parameters Eqs.~(\ref{eq:1.40}) and (\ref{eq:1.43}) assume
the form
\begin{equation}
  \label{eq:1.45}
  \dot{x}(\tau) = \sqrt{
  \frac{(n{-}2)\,x^\gamma {+}\gamma\,x^{2-n}}{n{-}2{+}\gamma}-1}\,,\qquad
  \ddot{x}(\tau) =  \frac{(n{-}2)\,\gamma}{2(n{-}2{+}\gamma)}
  \,\frac{(x^\gamma{-}x^{2-n})}{x}\,.
\end{equation}
At $x{=}1$ the system is in an equilibrium state with respect to the
variable $x(\tau)$ since according to Eqs.~(\ref{eq:1.45}), not only
$\dot{x}(\tau){=}0$, but also
$\ddot{x}(\tau){=}\dddot{x}(\tau){=}\dots{=}0$. (This is different
from the case without matter since in that case
$\ddot{x}(\tau)|_{x=1}{=}\gamma/2$ according to Eq.~(\ref{eq:1.42}).)
Like Einsteins solution for a static universe, this equilibrium is
unstable since $\ddot{x}{>}0$ for $x{>}1$ and $\ddot{x}{<}0$ for
$x{<}1$.%
\footnote{\label{foot:2} A similar but somewhat simpler solution of
  the cosmological equations, also starting from an unstable
  equilibrium, was published by this author in the year 2000
  \cite{Rebhan2}. As in Refs.~\cite{Vilenkin1} and~\cite{Vilenkin2},
  a constant density $\varrho_\Phi$, corresponding to a cosmological
  constant, was considered there; it was transfered into an initial
  equilibrium by the presence of matter as well.}%
~In contrast to the simpler cases mentioned in footnote~\ref{foot:2},
now also the dynamics of the field $\Phi$ must be determined from
Eqs.~(\ref{eq:0.2})-(\ref{eq:0.3}). In the following we restrict our
consideration to two special cases: 1.~validity of Eq.~(\ref{eq:1.42})
(no matter present), and 2.~validity of Eq.~(\ref{eq:1.45}) (matter
initially in equilibrium with the DE).  We can deal with both cases at
a time by introducing the definition
\begin{equation}
  \label{eq:1.46}
  g(x) = \left\{\begin{array}{cll}
      x^\gamma &\quad \mbox{for}&\quad\varrho_{m} \equiv 0\\
    \displaystyle \frac{(n{-}2)\,x^\gamma {+}\gamma\,x^{2-n}}{n{-}2{+}\gamma}&
    \quad \mbox{for}&\quad 
                      \displaystyle \varrho_{mi} = \frac{3\varrho_P}{8\pi}\frac{\gamma}
                      {(n{-}2{+}\gamma)}\,.
    \end{array} \right. 
\end{equation}
(Note that for $\varrho_{mi}{\neq}0$, this definition differs slightly
from that of Eq.~(\ref{eq:1.4}).) According to its definition, the
function $g(x)$ has the property
\begin{equation}
  \label{eq:1.47}
  g(1) = 1\,.
\end{equation}
With the above definition, the joint representation of Eqs.~~(\ref{eq:1.42})
and (\ref{eq:1.45}a) becomes
\begin{equation}
  \label{eq:1.48}
  \dot{x}(\tau) = \sqrt{g(x) -1}\,.
\end{equation}
\mbox{Using~$A\varrho'_\Phi(A){=}x\varrho'_\Phi(x){=}{-}(2{-}\gamma)\varrho_{\Phi
  i} x^{\gamma-2}\!$ and $\dot{\Phi}(\tau){=}t_P \dot{\Phi}(T)$
according to Eq.~(\ref{eq:1.37}b),} we obtain from
Eq.~(\ref{eq:1.20}b)%
\vspace*{-0.5\baselineskip}
\begin{equation} 
  \label{eq:1.49}
  \dot{\Phi}(\tau) = \frac{t_P c^2}{\hbar}\,
  \sqrt{\frac{(2{-}\gamma)\,\mu\,\varrho_{\Phi i}}{3}}\,x^{\gamma/2 -1}\,.
\end{equation}
With this, Eq.~(\ref{eq:1.38}a) and
$\dot{\Phi}(T){=}\dot{\Phi}(\tau)/t_p$ Eq.~(\ref{eq:0.2}a) yields
\begin{equation}
  \label{eq:1.50}
  V(x) = \frac{(4{+}\gamma)\,\varrho_{\Phi i}c^2}{6\,x^{2-\gamma}}\,,
\end{equation}
and using Eqs.~(\ref{eq:1.48}) and (\ref{eq:1.49}), we get
\begin{equation} 
  \label{eq:1.51}
  \frac{d\Phi}{dx} = \frac{\dot{\Phi}(\tau)}{\dot{x}(\tau)} =
  \frac{t_Pc^2}{\hbar}\,
  \sqrt{\frac{(2{-}\gamma)\,\mu\,\varrho_{\Phi i}}{3}}\,
  \frac{x^{\gamma/2}}{x\,\sqrt{g(x) -1}}\,.
\end{equation}
At $x{=}1$, according to Eqs.~(\ref{eq:1.49}) and (\ref{eq:1.51})
with~(\ref{eq:1.47}), we have $\dot{\Phi}(\tau){\neq}0$ and
$dx/d\Phi{=}0$. This means that the changes of $\Phi$ arising from
$\dot{\Phi}(\tau){\neq}0$ have no influence on the (unstable)
equilibrium $x{\equiv}1$ and $V(x){\equiv}V(1)$, or, otherwise,
equilibrium exist only with respect to $x$ and $V$, but not with
respect to $\Phi$ (see Fig.~\ref{fig:4}).

For non-equilibrium values $x$, integration of Eq.~(\ref{eq:1.51})
yields
\begin{equation}
  \label{eq:1.52}
  \Phi(x) = \frac{t_Pc^2}{\hbar}\,
  \sqrt{\frac{(2{-}\gamma)\,\mu\,\varrho_{\Phi i}}{3}}\,\int_{x_g}^x
  \frac{\xi^{\gamma/2}\,d\xi} {\xi\,\sqrt{g(\xi) -1}}\,.
\end{equation}
In the case $\varrho_m{\equiv}0$, in which there is no equilibrium at
$x{=}1$, we choose $x_g{=}1$ since the integral converges although the
integrand diverges like $1/\sqrt{\xi{-}1}$ for $\xi{\to}1$. In the
case described by Eq.~(\ref{eq:1.44}), the integrand behaves like
$1/(\xi{-}1)$ for $\xi{\to}1$ and is not integrable. In this case, we
choose $x_g{>}1$, e.g. $x_g{=}2$ in Fig.~\ref{fig:4}.

\subsubsection{Creation of the multiverse out of nothing by quantum
  tunneling}
\label{sec:quant-tunn}

According to Refs.~\cite{Vilenkin1} and~\cite{Vilenkin2}, the initial
state $x{=}1,\, \dot{x}(\tau){=}\,0$, start of the classical evolution
with zero expansion velocity, can be considered as coming about by
quantum-mechanical tunneling from nothing (no space, no time). An
approximate quasi-classical description of the tunneling process
preceding this classical high energy state can be obtained by
extending equation~(\ref{eq:1.48}) to the non-classical values $x{<}1$
yielding
\begin{displaymath}
  \dot{x}(\tau) = \mbox{i}\,\sqrt{1-g(x)}\,,
\end{displaymath}
and by employing an imaginary time
\begin{equation} 
  \label{eq:1.54}
  \tau = -\mbox{i}\,u \qquad\mbox{with~real}~u\,,
\end{equation}
a method introduced by Coleman~\cite{Coleman}. With this the equation
of motion assumes the ``Euclidean'' form%
\vspace*{-0.6\baselineskip}
\begin{equation}  
  \label{eq:1.55}
  \dot{x}(u) = \sqrt{1-g(x)}\,.
\end{equation}
For $g(x){=}x^\gamma$, we have $1{-}g(x){>}0$ in the quantum regime
$x{<}1$, and $\dot{x}(u)$ is well defined. The singularity of the
corresponding density $\varrho_{\Phi}{\sim}x^{\gamma-2}$ at $x{=}0$ is
tolerable because the total energy
$E{=}\varrho_{\Phi}c^2\mathcal{V}{=}2\pi c^2 L_P^3 \varrho_{\Phi
  i}x^{1+\gamma}$, the quantity which matters in the quantum regime,
vanishes as $x{\to}\,0$. By contrast in the case described by
Eq.~(\ref{eq:1.45}a), we would have $1{-}g(x){<}0$ for $x{<}1$, and the
connection to quantum tunneling becomes only possible, when a cutoff
of the density $\varrho_m$ is introduced. (This way simultaneously a
more serious singularity of $\varrho_m(x)$ at $x{=}0$ is
avoided). Accordingly, in the region $x{<}1$, we set%
\footnote{\label{foot:3}A cutoff at the Planck density could also be
  envisaged and would only marginally change the picture. However, the
  Plank density is a round figure only and has no precise physical
  explanation. The density $\varrho_i$ has the same order of
  magnitude and appears due to its physical background more
  appropriate for the cutoff.}%
\begin{equation}
  \label{eq:1.55b}
  \varrho_{m}\equiv \varrho_{m i} \qquad \mbox{and}\qquad 
  g(x) = \frac{(n{-}2)\,x^\gamma {+}\gamma}{n{-}2{+}\gamma}\,.
\end{equation}
At $x{=}1$, the solution of Eq.~(\ref{eq:1.55}) can be connected
continuously with the classical solution obtained from
Eq.~(\ref{eq:1.48}) for $x{\geq} 1$ by putting
\begin{equation}  
  \label{eq:1.56}
  \tau(x) = \int_1^x\frac{d\xi}{\sqrt{g(\xi)-1}}\,,\qquad
  u(x) = -\int_x^1\frac{d\xi}{\sqrt{1-g(\xi)}}\,.
\end{equation}
Inserting Eq.~(\ref{eq:1.54}) into the line element $ds^2$ yields
\begin{equation}
  \label{eq:1.57}
  ds^2 = - c^2\,du^2 - a^2(u)\,dr^2 + \dots\,,
\end{equation}
referred to as Euclidean form. This shows that the introduction of an
imaginary time amounts to converting $ct$ it into a fourth space
coordinate $cu$.

\begin{figure}
  \includegraphics[width=0.47\textwidth]{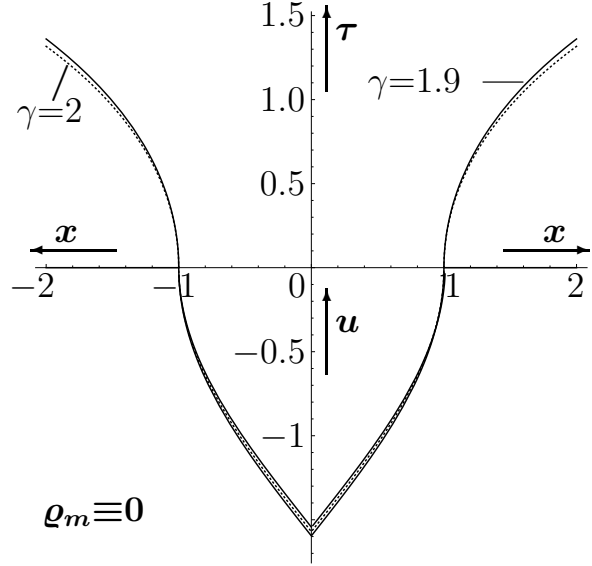} 
  \caption{\label{fig:1} \small Normalized age of the multiverse, $\tau$ for
    $x{\geq}1$ and $u{=}\mbox{i}\tau$ for $0{\leq}x{<}1$, as function
    of the relative cosmic scale factor $x{=}A/l_P$ in the case
    $\varrho_{m}{\equiv}0$. The mirror images of the curves $\tau(x)$
    and $u(x)$ on the left-hand side express the fact that at every
    time $\tau$, the projection of the spherically symmetric boundary
    of the multiverse onto the $x,\tau$-plane yields two points
    symmetrically positioned with respect to the $\tau$-axis.  The
    dotted curve obtained for $\gamma{=}2$ represents de Sitter
    space. In the quantum regime $0{\leq}x{<}1$, the upper solid curve
    is obtained when the divergent density
    $\varrho_\Phi{=}\varrho_{\Phi i}x^{\gamma-2}$ is truncated by
    replacing it with $\varrho_\Phi{\equiv}\varrho_{\Phi i}$.}
\end{figure}

\begin{figure}
  \includegraphics[width=0.40\textwidth]{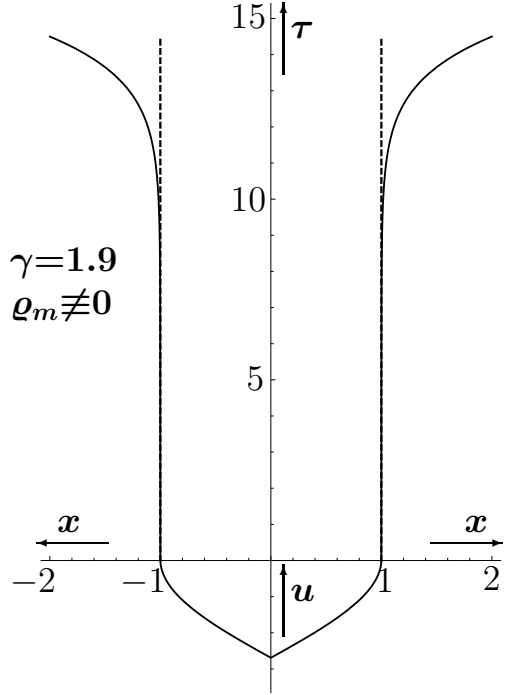}
  \caption{\label{fig:2} Curves $\tau(x)$ and $u(x)$ in the case of
    simultaneous tunneling of matter, $\varrho_{m i}$ and
    $\varrho_{\Phi i}$ chosen according to Eqs.~(\ref{eq:1.44}). The
    classical evolution, represented by the upper solid curves, starts
    at $\tau{=}0$ from a small perturbation of the equilibrium state
    represented by the dashed vertical lines.}
\end{figure}

\begin{figure}
  \includegraphics[width=0.58\textwidth]{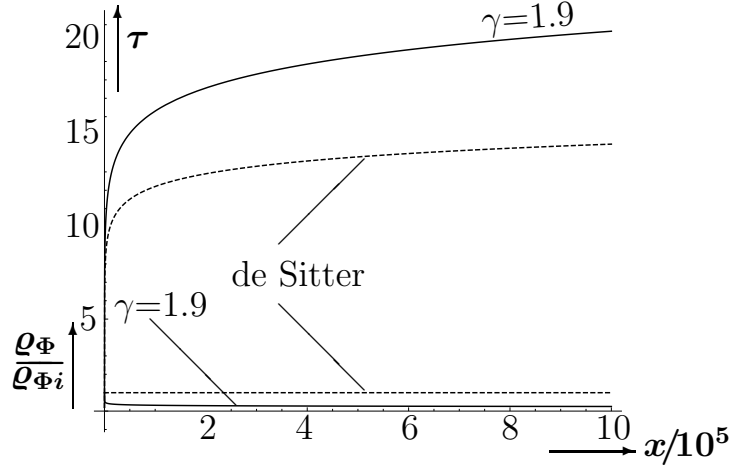}
  \caption{\label{fig:3} Normalized age $\tau(x)$ and normalized mass
    density $\varrho_{\Phi}(x)/\varrho_{\Phi i}$ of the multiverse for
    larger values of $x$, the solid curves representing
    $\gamma{=}1.9, \varrho_m{\equiv}0$ and the dotted ones de Sitter
    space ($\gamma{=}2$). (The curves are obtained from
    Eq.~(\ref{eq:1.38}a) and Eqs~(\ref{eq:1.56}a).)}
\end{figure}

In Figs.~\ref{fig:1} and~\ref{fig:2}, it is shown for
$\varrho_m{\equiv}0$, $\gamma{=}1.9$ and for $\varrho_m{\not\equiv}0$,
$n{=}3$, $\gamma{=}1.9$, respectively, how the expansion parameter $x$
evolves from $x{=}0$ in the quantum regime $x{<}1$ to values $x{>}
1$. The tip at the bottom of the figures does not conform with the no
boundary proposal of Hartle and Hawking~\cite{Hartle-Hawking} (nicely
illustrated in Ref.~\cite{Jorma_Louko}); on the other hand, our
approach agrees widely with that of Vilenkin (see
Refs.~\cite{Vilenkin1} and~\cite{Vilenkin2}).%
\footnote{The picture obtained here (Fig.~1) is widely similar to that
  of Fig.~1(a) in Ref.~\cite{Vilenkin1}. Only the semicircle at the
  bottom of Vilenkin's figure does not exhibit a tip at $x{=}0$ as our
  figure what it actually should: the inverse
  $t(a){=}H^{-1}\arccos(Ha)$ of Vilenkin's solution
  $a(t){=}H^{-1}\cos(Ht)$, which can easily be obtained by putting
  $\gamma{=}2$ in Eq.~(\ref{eq:1.55}) with~(\ref{eq:1.46}a) (dotted
  curve in Fig.~\ref{fig:1}), has a tip there. Note, however, that
  according to Vilenkin his figure is only a symbolical
  representation.}%
~In Fig.~\ref{fig:3}, a further comparison of the case
$\gamma{=}1.9, \varrho_{mi}{\equiv}0$ with de Sitter space shows that
for larger values of $x$, the differences between the two become
appreciable.

In the case $\varrho_\Phi{=}const$ considered in
Ref.~\cite{Vilenkin1}, Eqs.~(\ref{eq:0.2})-(\ref{eq:0.3}) are
trivially fulfilled with $\Phi{=}const$ and $V(\Phi){=}const$, while
in our case, their solution becomes nontrivial. Introducing the new
time parameter $u{=}\mbox{i}\tau$ in Eq.~(\ref{eq:1.49}) reveals that
owing to
$d\Phi/du{=}(d\Phi/d\tau)\,(d\tau/du){=}{-}\mbox{i}\,d\Phi/d\tau$, the
quantity $\Phi$ becomes imaginary whence we set
\begin{displaymath}
  \Phi\big(\tau(u)\big) = -\mbox{i}\,\varphi(u) 
  \qquad\mbox{with~real}\qquad \varphi(u)\,.
\end{displaymath}
With this and Eq.~(\ref{eq:1.54}), Eq.~(\ref{eq:1.49}) is converted into
\begin{displaymath}
  \dot{\varphi}(u) = \frac{t_P c^2}{\hbar}\,
  \sqrt{\frac{(2{-}\gamma)\,\mu\,\varrho_{\Phi i}}{3}}\,x^{\gamma/2 -1}\,.
\end{displaymath}
From this and Eq.~(\ref{eq:1.55}) in analogy to Eq.~(\ref{eq:1.52}) we
obtain
\begin{displaymath}
  \varphi(x) = -\frac{t_Pc^2}{\hbar}\,
  \sqrt{\frac{(2{-}\gamma)\,\mu\,\varrho_{\Phi i}}{3}}\,\int_x^1
  \frac{\xi^{\gamma/2}\,d\xi} {\xi\,\sqrt{1-g(\xi)}} \,.
\end{displaymath}
Using%
\vspace*{-\baselineskip}
\begin{displaymath}
  \varphi(0) = -\frac{t_Pc^2}{\hbar}\,
  \sqrt{\frac{(2{-}\gamma)\,\mu\,\varrho_{\Phi i}}{3}}\,J_\gamma
  \qquad\mbox{with}\qquad
  J_\gamma = \int_0^1 \frac{\xi^{\gamma/2}\,d\xi} {\xi\,\sqrt{1-g(\xi)}}
\end{displaymath}
we can bring  this and the result~(\ref{eq:1.52}) into the forms
\begin{equation}
  \label{eq:1.58}
  \tilde{\varphi}(x) = \frac{\varphi(x)}{|\varphi(0)|} 
  = -\frac{1}{J_\gamma}\int_x^1\!
  \frac{\xi^{\gamma/2}\,d\xi} {\xi\,\sqrt{1-g(\xi)}}\,,
  \quad
  \tilde{\Phi}(x) = \frac{\Phi(x)}{|\varphi(0)|}  
  = \frac{1}{J_\gamma}\int_{x_g}^x
   \!\frac{\xi^{\gamma/2}\,d\xi} {\xi\,\sqrt{g(\xi)-1}}\,,
\end{equation}
and Eq.~(\ref{eq:1.50}) can be written as
\begin{equation}
  \label{eq:1.59}
  \tilde{V}(x) = \frac{V(x)}{V_i} = x^{\gamma-2}
  \qquad\mbox{where}\qquad 
  V_i = V(1)= \frac{(4{+}\gamma)\,\varrho_{\Phi i}c^2}{6}\,.
\end{equation}
\begin{figure}
  \includegraphics[width=0.58\textwidth]{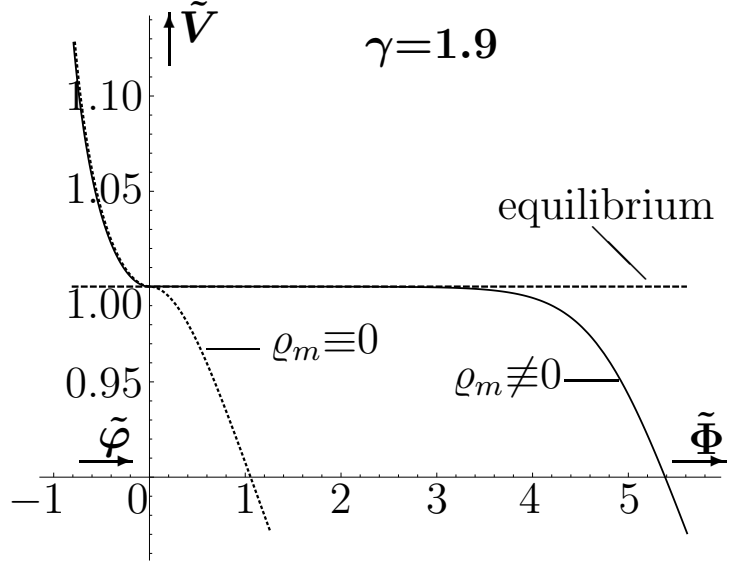} 
  \caption{\label{fig:4} $\tilde{V}(\tilde{\varphi})$ is the
    normalized potential $V$ in the quantum regime (to the left of the
    $\tilde{V}$-axis), and $\tilde{V}(\tilde{\Phi})$ is the potential
    in the classical regime (to the right of the $\tilde{V}$-axis),
    shown for the case without matter (dotted curves) and with matter
    (solid curves), $\varrho_{m i}$ being chosen according to
    Eq.~(\ref{eq:1.44}a). On the dashed horizontal line, matter and
    $\Phi$-field are in an unstable equilibrium. For the solid curve
    on the right representing a deviation from equilibrium, we have
    chosen $n{=}3$ in Eq.~(\ref{eq:1.44}a), $x_g{=}2$ in
    Eq.~(\ref{eq:1.58}b), $\tilde{V}$ according to
    Eq.~(\ref{eq:1.59}a) (slightly below $1$) at an $x$ slightly above
    $1$, and added an integration constant on the right hand side of
    Eq.~(\ref{eq:1.58}b) such that $\tilde{\varphi}$ and
    $\tilde{\Phi}$ connect continuously at their respective zero.}
\end{figure}
Eqs.~(\ref{eq:1.58}) and (\ref{eq:1.59}) constitute parametric
representations of the functions~$\tilde{V}(\tilde{\varphi})$ and
$\tilde{V}(\tilde{\Phi})$ shown in Fig.~\ref{fig:4} for the cases
with and without matter. When in the latter case, the ansatz
$\varrho_\Phi{\sim}x^{\gamma-2}$ is retained within the tunneling
regime $x{<}1$, then $\tilde{V}(\tilde{\varphi})$ diverges together
with $\varrho_\Phi$ at $x{=}0$, and the shape of the combined curves
$\tilde{V}(\tilde{\varphi})$ and $\tilde{V}(\tilde{\Phi})$ resembles
that of a ski jump with vanishing slope at $x{=}0$. Introducing a
cutoff by keeping $\varrho_\Phi$ at the value $\varrho_{\Phi i}$ in
the whole tunneling regime leads to $V(\varphi){=}V_i{=}const$ or
$\tilde{V}(\tilde{\varphi}){=}1$, respectively (dashed curve). In the
case $\varrho_{m i}{\neq}0$, regularized according to
Eqs.~(\ref{eq:1.55b}) in the quantum regime, the form of the curve
$\tilde{V}(\tilde{\varphi})$ is almost exactly the same as without
matter there, while the perturbational solution
$\tilde{V}(\tilde{\Phi})$ in the classical regime remains much longer
in the neighborhood of the (unstable) equilibrium.

\section{Taking into account subuniverses}
\label{sec:account_sub}

\subsection{Relation between mass densities in our universe and its
  associated multiverse}
\label{sec:Relations}

For identification of the DE presently observed in our universe with
the DE of its associated multiverse, represented by the field $\Phi$,
we must determine how the mass densities of the two are
related. Within the spatial dimensions of our universe or comparable
subuniverses (coordinate system $S_s$ with time $t$, scale factor $a$
and mass density $\varrho_{s\Phi}$ of the field $\Phi$), which in
relation to the associated multiverse are very small, the effects of
the spatial curvature are negligible (see Eq.~(\ref{eq:1.34a}))
so that the square of the line element can be written as
\begin{equation}
  \label{eq:2.1}
  ds_s^2 = c^2\,dt^2 -a^2(t)\,dr^2 - a^2(t)\,r^2\,d\Omega
  \qquad\mbox{with}\qquad
  d\Omega = d\vartheta^2+\sin^2\vartheta\,d\varphi^2\,.
\end{equation}
In the $\Phi$-multiverse (system $S_M$) we have
\begin{equation}
  \label{eq:2.2}
  ds_M^2 = c^2\,dT^2 -A^2(T)\,d\chi^2 - A^2(T)\,\sin^2\chi \,d\Omega'
  \qquad\mbox{with}\qquad
  d\Omega' = d\vartheta'^2+\sin^2\vartheta'\,d\varphi'^2\,.
\end{equation}
The main question is how the density $\varrho_\Phi$ of the
$\Phi$-multiverse enters the equations valid in our universe and vice
versa. The usual general relativistic equations for the transformation
between different coordinate systems refer to the same physical
situation. We construct such a situation by conceptually removing the
matter density $\varrho_{sm}$ from our universe, this way transferring
the total mass density $\varrho_{sm}{+}\varrho_{s\Phi}$ into
$\varrho_{s\Phi}$; simultaneously we maintain the coordinates
$t,r,\vartheta$ and $\varphi$ of the universe. Now, we consider the
same situation from the coordinate system $T,\chi,\vartheta'$ and
$\varphi'$ of the $\Phi$-multiverse where
$\varrho_\Phi{=}\varrho_\Phi(T)$. Thereby we assume that the origins
of the two systems coincide, whence for symmetry reasons the angles
$\vartheta$ and $\varphi$ can also be arranged to coincide with
$\vartheta'$ and $\varphi'$ respectively, i.e.
\begin{equation}
  \label{eq:2.3}
  \vartheta' = \vartheta\,,\qquad \varphi' = \varphi\,.
\end{equation}
Thus, the mass density $\varrho_{\Phi}(T)$ of the $\Phi$-multiverse will in
general appear as $\varrho_{s\Phi}(t,r)$ in our universe.

The mass density $\varrho$ is a constituent of the energy-momentum
tensor $T_{\mu\nu}{=}\varrho\,U_{\mu}U_{\nu}$ and therefore transforms
according to
\begin{displaymath}
  \varrho'\,U'_{\mu}U'_{\nu} = T'_{\mu\nu} = T_{\alpha\beta}\,
\frac{\partial x^\alpha}{\partial x'^\mu}\,\frac{\partial x^\beta}{\partial x'^\nu}
   = \varrho\,U_{\alpha}U_{\beta}\,\frac{\partial x^\alpha}{\partial x'^\mu}\,
   \frac{\partial x^\beta}{\partial x'^\nu}\,.
\end{displaymath}
Considering $x^\alpha$ as the coordinates of $S_s$ (our universe) and
$x'^\mu$ as those of $S_M$, and identifying $\varrho$ and $\varrho'$
with the rest densities (whence $U_0{=}U'_0{=}c$ and
$U_k{=}U'_k{=}0$ for $k{=}1,2,3$), from this we get
\vspace*{-0.8\baselineskip}
\begin{displaymath}
 \varrho_{s\Phi} =  
 \varrho_{\Phi} \Big/\!\left(\frac{\partial t}{\partial T}\right)^2\,.
\end{displaymath}
$\varrho_{s\Phi}$ must depend on $t$ only and does it, when
$T{=}T(t)$. On the special assumption 
\begin{equation}
  \label{eq:2.4}
  T = T_{si} + t \,,
\end{equation}
with $T_{si}{=}$ time of creation of the subuniverse, we finally have
\begin{equation}
  \label{eq:2.5}
  \varrho_{s\Phi}(t) = \varrho_\Phi(T)
  \qquad\mbox{for}\qquad
   T = T_{si} + t \,.
\end{equation}
According to Eqs.~(\ref{eq:2.1}) and (\ref{eq:2.2}), $t$ and $T$ are
metric times measured with identical standard
clocks. Eq.~(\ref{eq:2.5}) exhibits the space independence of
$\varrho_{s\Phi}$ demanded in requirement~\ref{i3} of the
Introduction.

\subsection{Influence of the associated multiverse on the 
  evolution  of our universe}
\label{sec:infl-sub}

According to our model,  $\Phi$-multiverses provide the background for
the creation of subuniverses which for their part evolve from
appropriate fluctuations of independent inflatons physically different
from $\Phi$. Now, we turn to the question how $\Phi$ affects the
evolution of subuniverses, considering our universe as a typical
example. Already, shortly after its creation, the primordial matter
content of the multiverse is so much diluted that, as in
Section~\ref{sec:late-ev}, it can be completely neglected. During the
inflation of our universe, also $\varrho_{s\Phi}$ is so much smaller
than the mass density of its inflaton, that it can be neglected as
well, and according to table~\ref{tab:1}, inside the universe, the
curvature of the multiverse can be omitted all the more. In
consequence, the early evolution of our universe proceeds according to
usual concepts, must not be reviewed here, and we can restrict our
consideration to the period after inflation and photon decoupling. In
this era, matter obeys the equation
$\varrho_{sm}{=}\varrho_{sm 0}\,a_0^3/a^3(t)$. The influence of the
$\Phi$-multiverse on our universe consists in imprinting~$\Phi$ on it
as an external field with prescribed dynamics given by
Eq.~(\ref{eq:1.35}a) together with Eq.~(\ref{eq:1.32}). With neglect
of the curvature term ${-}c^2a^2/A^2$, the Friedmann-Lemaitre equation
of our universe becomes
\begin{displaymath} 
  \dot{a}^2(t) = \frac{8\pi G}{3}\,(\varrho_{sm}+\varrho_{\phi})\,a^2 
  = \frac{8\pi G}{3}\,\left(\frac{\varrho_{sm 0}\,a_0^3}{a}
  +\varrho_{\Phi 0}\,x^{\gamma-2}a^2\right)
\end{displaymath}
or%
\vspace*{-0.5\baselineskip}
\begin{displaymath} 
  \dot{y}^2(t)   = \frac{8\pi G}{3}\,\left(\frac{\varrho_{sm 0}}{y}
    +\varrho_{\Phi 0}\,x^{\gamma-2}y^2\right)
  \qquad\mbox{with}\qquad
  y = \frac{a}{a_0}\,.
\end{displaymath}
For calculating the acceleration $\ddot{y}(t)$ at the present time,
$t_0$ in our universe and $T_0{=}T_{si}+t_0$ in the multiverse, we
differentiate the last equation with respect to $t$ and then divide it
by $2\dot{y}(t)$. Using the approximation
\begin{displaymath}
  \frac{dx}{dt} = \dot{x}(\tau)\,\frac{d\tau}{dT}\frac{dT}{dt}
   =\frac{0.826\,x^{\gamma/2}}{t_{H 0}} = 0.826\,H_0\,x^{\gamma/2}\,,
\end{displaymath}
following from Eqs. (\ref{eq:1.27}), (\ref{eq:1.1}a), (\ref{eq:1.2})
and~(\ref{eq:2.4}), furthermore
$\dot{y}_0{=}\dot{y}(t_0){=}\dot{a}(t_0)/a_0{=}H_0$, $x_0{=}y_0{=}1$
and
$H_0^2{=}8\pi G(\varrho_{sm0}{+}\varrho_{\Phi 0})/3{=}8\pi
G\varrho_{c0}/3$, and finally using
$\varrho_{sm 0}{=}0.317\,\varrho_{c0}$ and Eq.~(\ref{eq:1.5}), after
some rearrangement, we obtain
\begin{equation}
  \label{eq:2.6}
  \ddot{y}_0 = \ddot{y}(t_0) = H_0^2\,
  \Big[0.525-0.282(2{-}\gamma)\Big]\,.
\end{equation}
Positive acceleration is achieved for $\gamma {>}\, 0.14$. For later
purposes also the present values of $\dddot{y}$ and the time
derivative of the deceleration parameter
$q{=}{-}\ddot{a}/(H^2a){=}{-}\ddot{y}\,y/\dot{y}²$ (present value
$q_0{=}{-}\ddot{y}_0/H_0$) are noted down, which can be derived in a
similar way:
\begin{equation}
  \label{eq:2.7}
  \dddot{y}_0 = H_0^3\,\Big[1-(2{-}\gamma)(0.441{+}0.270\gamma)\Big]\,,
  \quad
  \dot{q}_0 = -H_0\Big[9.74-(2{-}\gamma)(0.449{+}0.111\gamma)\Big]\,.
\end{equation}

In table~\ref{tab:1} for several values of $\gamma$, the acceleration
$\ddot{y}_0$ is entered in multiples of the acceleration%
\vspace*{-0.5\baselineskip}
\begin{displaymath}
  \ddot{y}_\Lambda = \frac{\ddot{a}_\Lambda }{a_0} =
  \frac{4\pi G}{3}\,(2\,\varrho_{\Lambda} -\varrho_{sm 0})
  = 0.525\,H_0^2  
\end{displaymath}
induced by a cosmological constant $\Lambda$ corresponding to a mass
density \mbox{$\varrho_\Lambda{=}\varrho_{\Phi 0}$}. In order that the
present acceleration by the field $\Phi$ is comparable with that by
$\Lambda$, we must have $\gamma{\gtrsim}1.95$ and correspondingly high
values of $\zeta$ according to table~\ref{tab:1}.  In units of
$t_{H0}$, the age of our universe is $\tau_u{=}0.98$. With this and
Eq.~(\ref{eq:1.31}b), we get from Eq.~(\ref{eq:1.32}) for the
expansion $x_s$ of the multiverse at the start of our universe
\begin{displaymath}
  x_s = \left(\frac{\tau_s}{\tau_0}\right)^{\!\frac{1}{1-\gamma/2}}\!
  = \left(\frac{\tau_0-0.98}{\tau_0}\right)^{\!\frac{1}{1-\gamma/2}}\!
  = \left(1-\frac{0.813}{0.83\,\tau_0}\right)^{\!0.83\,\tau_0}\!\!
    < \, \mbox{e}^{-0.813} \approx 0.443\,.
\end{displaymath}
This is listed in table~\ref{tab:1} for several values of $\gamma$ and
shows that during the lifetime of our universe, the expansion of the
multiverse more than doubles. For the density $\varrho_{\Phi h}$ at
half of its lifetime, $\tau_h{=}\tau_0{-}0.49$, from
Eqs.~(\ref{eq:1.25}a) and (\ref{eq:1.32}), we get
\begin{displaymath}
  \frac{\varrho_{\Phi h}}{\varrho_{\Phi 0}} = x_h^{\gamma-2} = 
  \left(\frac{\tau_h}{\tau_0}\right)^{\frac{\gamma-2}{1-\gamma/2}} =
 \left(\frac{\tau_0}{\tau_0-0.49}\right)^2\,.
\end{displaymath}
This is again shown in table~\ref{tab:1}; for $\gamma{\gtrsim}1.95$, the
density $\varrho_{\Phi}$ changes so little that its action on the
dynamics of our universe is almost the same as that of a cosmological
constant. In particular, it causes the same acceleration of the present
expansion, and the change from deceleration to acceleration occurs at
about half the life time of our universe.

\subsection{Influence of the subuniverses on the evolution 
  of the multiverse}
\label{sec:accel-exp}

So far, the evolution of the multiverse was treated without regard to
its subuniverses. To make up for this, we determine the velocity at
which our universe expands into its associated multiverse. Since we
are only interested in the bulk motion, we employ a very simple model
of the boundary between the two, assuming a discontinuous jump of
$\varrho_{sm}$ from a uniform inside value to zero
outside. Furthermore, we assume again that origin and angles of the
spatial coordinates of multiverse and universe coincide.

In the coordinate system $S_s$ of our universe, the position of the
boundary is $r{=}r_b{=}const$. For a fixed point on it,
$dr{=}d\vartheta{=}d\varphi{=}0$ whence from Eq.~(\ref{eq:2.1}) we
get
\begin{displaymath}
  ds_s^2 = c^2\,dt^2\,.
\end{displaymath}
In the system $S_M$ of the multiverse, the position $\chi_b(T)$ of this
point will in general change with $T$, whence with
$d\vartheta{=}d\varphi{=}0$, we obtain from Eq.~(\ref{eq:2.2})
\begin{displaymath}
  ds_M^2 = c^2\,dT^2-A^2(T)\,d{\chi_b}^2 = 
  [c^2-A^2(T)\,\dot{\chi}_b^2(T)]\,dT^2\,.
\end{displaymath}
The invariance $ds_s{=}ds_M$ of the line element with respect to coordinate
transformations leads to%
\vspace*{-0.5\baselineskip}
\begin{equation}
  \label{eq:2.8}
  \frac{dt}{dT} = \sqrt{1-v^2/c^2} 
  \qquad\mbox{with}\qquad
  v = A(T)\,\dot{\chi}_b(T)\,.
\end{equation}
The following calculations serve for the evaluation of $v$ or
$\dot{\chi}_b(T)$ respectively. For the distance $dl$ of two
neighboring points on the boundary $r{=}r_b$ which differ only
in~$\vartheta$ (i.e. $dt{=}0$ and $dr{=}d\varphi{=}0$), we get from
Eq.~(\ref{eq:2.1})
$dl_s^2 {=}{-}ds_s^2 {=} a^2(t)\,r_b^2\,d\vartheta^2$ in $S_s$, and
from Eqs.~(\ref{eq:2.2})-(\ref{eq:2.3}) we get
$dl_M^2 {=}{-}ds_M^2{=}A^2(T)\,\sin^2\chi_b\,d\vartheta^2$ in $S_M$,
because according to Eq.~(\ref{eq:2.8}) or
$dT{=}dt/\sqrt{1{-}v^2/c^2}$ we have $dT{=}0$ for $dt{=}0$ and
$d\chi_b{=}\dot{\chi}_b(T)\,dT{=}\,0$. In this case the invariance of
the line element, $dl_s{=}ds_M$, leads to%
\vspace*{-0.5\baselineskip}
\begin{equation}
  \label{eq:2.9}
  a(t)\,r_b = A(T)\,\chi_b(T)\,,
\end{equation}
where $A\,\sin\chi_b$ was replaced by $A\,\chi_b$ since
$\sin\chi_b{=}r_b\,a/A$ is extremely small.  Differentiating
Eq.~(\ref{eq:2.9}) with respect to $T$ and then inserting
Eqs.~(\ref{eq:2.8}) leads to
\begin{equation}
  \label{eq:2.10}
  \dot{a}(t)\,r_b\,\sqrt{1{-}v^2/c^2} = \dot{A}(T)\,\chi_b + v
\end{equation}
and resolving this equation with respect to $v$ eventually yields
\begin{equation}
  \label{eq:2.11}
  v = \frac{\dot{a}(t)\,r_b\,\sqrt{1{+}(\dot{a}(t)\,r_b/c)^2
      {-}(\dot{A}(T)\,\chi_b/c)^2}-\dot{A}(T)\,\chi_b}
  {1+(\dot{a}(t)\,r_b/c)^2}\,.
\end{equation}
(A second solution with a minus sign in front of the root term was
discarded since for $a(t){\to}\,0$ or equivalently $\chi_b{\to}\,0$ (see
Eq.~(\ref{eq:2.9})), Eq.~(\ref{eq:2.10}) requires
\mbox{$v\,{\to}\,\dot{a}(t)\,r_b\sqrt{1{-}v^2/c^2}>0$.)} For evaluating $v$ at
the present time $t_0$, we use
\begin{equation}
  \label{eq:2.12}
  \dot{a}(t_0)\,r_b =H_0\,a_0\,r_b 
  = \frac{a_0\,r_b}{t_{H0}} = \frac{2R}{t_{H0}}
  = 3.36\,c\,.
\end{equation}
Thereby, for the present distance of the outer boundary of the universe,
\begin{equation}
  \label{eq:2.13}
  a_0\,r_b = 2R
\end{equation}
was assumed, so that inhomogeneities propagating from outer regions
inwards cannot have spoiled the homogeneity and isotropy inside the
observable universe $a_0r{\leq}R$ (see p.~231 of
Ref.~\cite{Mukhanov}); furthermore, Eqs.~(\ref{eq:1.2}) and
(\ref{eq:1.7}) were used. From Eqs.~(\ref{eq:1.1}), with the
approximation \mbox{$\dot{x}(\tau){=}0.826\,x^{\gamma/2}$} of
Eq.~(\ref{eq:1.27}), and with $x{=}1$ we get
\begin{equation}
  \label{eq:2.13a}
  \dot{A}(T_0)= \left.A_0\,\dot{x}(\tau)\frac{d\tau}{dT}\right|_{T_0}
  = \frac{0.826\,A_0}{t_{H0}}\,,
\end{equation}
and from Eq.~(\ref{eq:2.13}) with Eqs.~(\ref{eq:2.9}) and
(\ref{eq:1.13}b), we obtain
\begin{equation}
  \label{eq:2.14}
  2R = A_0\,\chi_{b0} = \zeta\,R\,\chi_{b0}
  \qquad\mbox{or}\qquad
  \chi_{b0} = \frac{2}{\zeta}
\end{equation}
where $\chi_{b0}{=}\chi_b(T_0)$. Combining Eqs.~(\ref{eq:2.13a})
and~(\ref{eq:2.14}a), and using Eqs.~(\ref{eq:1.2}) and~(\ref{eq:1.7})
yields%
\begin{displaymath}
  \dot{A}(T_0)\,\chi_{b0} = \frac{0.826
    \, A_0\,\chi_{b0}}{t_{H0}}
  = \frac{1.65\,R}{t_{H0}} = 2.77\,c\,.
\end{displaymath}
Inserting this and the result~(\ref{eq:2.12}) in
Eqs.~(\ref{eq:2.11}) and (\ref{eq:2.8}a), we finally obtain
\begin{equation}
  \label{eq:2.15}
  v(T_0) = 0.36\,c\,,
  \qquad  
  \left.\frac{dt}{dT}\right|_{t_0} = 0.93\,,
\end{equation}
and
\vspace*{-0.5\baselineskip}
\begin{equation}
  \label{eq:2.16}  
  \dot{\chi}_b(T_0) = \frac{v(T_0)}{A_0}=\frac{0.36\,c}{\zeta\,R} 
  = \frac{0.48\cdot 10^{-18}\,\mbox{s}^{-1}}{\zeta}\,.
\end{equation}
According to Eq.~(\ref{eq:2.14}b) the present value of $\chi_{b0}$ is
extremely small since according to table~\ref{tab:1}, $\zeta$ is
extremely large for $\gamma{\gtrsim}1.95$. At the present rate of
change, the time required for doubling $\chi_{b0}{=}2/\zeta$ is%
\vspace*{-0.5\baselineskip}
\begin{displaymath}
  \Delta T \approx \frac{\Delta \chi}{\dot{\chi}(T_0)}=\frac{2/\zeta}
  {0.48\cdot 10^{-18}\,\mbox{s}^{-1}/\zeta} = 4.17\cdot 10^{18}\,\mbox{s}
  = 9.5\,t_{H0}\,.
\end{displaymath}
After this time, the matter density $\varrho_{sm}$ of our universe has
already for a long time become so small that it can completely be
neglected. In consequence, the volume occupied by it can be treated
like the empty regions of the multiverse and expand like these as
treated in Section~\ref{sec:Evo without}.

The regions of the multiverse not occupied by subuniverses are not
influenced by the latter, because the gravitational field created by
them cannot extend beyond their boundary: This is essentially due to
Birkhoff's theorem (time-independence of all metric coefficients in
the vacuum surrounding a spherically symmetric mass or energy
distribution, see Ref.~\cite{Birkhoff} or, e.g.
Ref.~\cite{Weinberg2}, in combination with the fact that the
gravitational field of the subuniverse was zero outside of it before
its emergence).%
\footnote{Deviating from the requirements of Birkhoff's theorem, the
  universe is not surrounded by vacuum but by the field $\Phi$. The
  theorem is valid, however, when in a first step, the field $\Phi$ is
  removed from the surroundings. Nothing changes within the universe
  when the field is brought back to them in a second step, because a
  homogeneous mass distribution exhibits no gravitational field within
  a spherical cavity. (Homogeneous mass distribution can be assumed on
  large scales since the probability for the emergence of subuniverses
  can be assumed to be space- and time-independent.)  Since there is
  no action on the universe, there is no reaction on the
  surroundings.}%
~In consequence the unoccupied regions evolve like the corresponding
regions of a $\Phi$-multiverse, which is permanently devoid of matter
as treated in Section~\ref{sec:Evo without}.

According to what was said above, after some time, the regions occupied
by subuniverses behave essentially like unoccupied regions. Therefore,
the evolution of the all-embracing space-time can be treated without
regard to subuniverses for all times. (We exclude the possibility
that the birth rate of subuniverses would be so high and pack them so
densely that they would collide and merge in a stadium when their
density cannot yet be neglected.)

\section{Discussion of cosmological problems, initial
  conditions and observational constraints}
\label{sec:obs-const}

It can be taken from table~\ref{tab:1} that for all parameter values
$\gamma$, the (normalized) radius $\zeta{=}A/R$ of the multiverse is
very large. According to Eqs.~(\ref{eq:1.33}) and (\ref{eq:1.34}),
$\zeta$ and $\zeta_m$ grow exponentially with the present (normalized)
age $\tau_0$ of the multiverse. Already, at the low age
$\tau_0{=}3.2$, we have $\zeta{=}10^{100}$ and an accordingly small
curvature. Thus, for practically all values~$\gamma$, the curvature of
the space which our universe lives in, is immeasurably small. This
means that the flatness problem of our universe resolves quite
naturally by simply not existing.

The phenomenon of dark energy involves two puzzling problems: Why is
its presently observed density $\varrho_{\Phi 0}$ so small, and why
does it almost coincide with the present density $\varrho_{m 0}$ of
matter? To examine these questions within the scope of the current
model, using Eqs.~(\ref{eq:1.5}), (\ref{eq:1.14}a), (\ref{eq:1.16}),
(\ref{eq:1.17}) and~(\ref{eq:1.44}b), and restricting ourselves to the
cases $\varrho_{m i}{=}0$ and $\varrho_{m i}$ given by
Eq.~(\ref{eq:1.44}a), at first, we calculate
\begin{displaymath}
   \varrho_{\Phi i} =\left\{
     \begin{array}{lll}
       \varrho_{i} \quad&\mbox{for}\quad&\varrho_{mi}{=}0\\
       \varrho_{i}/(1{+}\gamma)\quad&\mbox{for}\quad&
       \varrho_{mi}{\neq}0\,,\;n{=}3\\
       2\,\varrho_{i}/(2{+}\gamma)\quad&\mbox{for}\quad&
       \varrho_{mi}{\neq}0\,,\;n{=}4
     \end{array}
     \right\}
     \qquad\mbox{with}\quad
  \varrho_{i} = 0.98\cdot 10^{122}\,\varrho_{\Phi 0}\,.
\end{displaymath}
This holds for any multiverse created out of nothing and shows that
initially $\varrho_\Phi$ has roughly the value following for quantum
fluctuations from elementary particle physics with cut-off, but
without renormalization and regard to symmetry breaking (see
e.g. Ref.~\cite{Weinberg}). It may be pointed out here, that the
initial condition~(\ref{eq:1.8}), which is responsible for the
creation out of nothing, is not only satisfied by the initial
values~(\ref{eq:1.13}a) and~(\ref{eq:1.16}), but also by
$x_i{=}\lambda\,l_P/A_0$ and
$\varrho_{i}{=}3\varrho_P/(8\pi\lambda^2)$ with $\lambda{>}1$. This
provides some flexibility which might be useful for adjustments
concerning the initial magnitude of the multiverse or the initial
densities.

Subuniverses of earlier origin than ours had to cope with higher
values $\varrho_{\Phi}{>}\varrho_{\Phi 0}$. In order that the galaxies
of our universe could develop in such a way as we observe them,
$\varrho_\Phi$ cannot be much larger than $\varrho_{\Phi 0}$, a factor
10 would already be too much. Thus, against higher values, the
anthropic principle can be invoked. Much smaller values would,
however, still be possible without much change in the appearance of
our universe, of course, except for the observed acceleration of the
expansion. In the framework of a multiverse with different generations
of subuniverses, the present value
$\varrho_{\Phi 0}{=}0.683\varrho_{c 0}$ in our universe must in view
of possible smaller values be considered as a matter of chance. On the
other hand, the approximate coincidence with $\varrho_{m 0}$ is a
consequence of this, since in a curved universe with practically
vanishing spatial curvature we have
$\varrho_{m 0}{=}\varrho_{c 0}{-}\varrho_{\Phi 0}{=}0.317\varrho_{c 0}
{=}0.464\varrho_{\Phi 0}$.

The ratio $w\,{\equiv}\,p_\Phi/(c^2\varrho_\Phi)$ is usually
considered as the most important quantity for the characterization of
DE properties. Modeling DE by a cosmological constant yields
$w{=}{-}1$. Time dependent models of DE like quintessence usually lead
to a time dependent $w(t)$ which is frequently approximated by a
constant value. From Eqs.~(\ref{eq:0.2}), (\ref{eq:1.25}a) and
(\ref{eq:1.36}a) for our model we get%
\vspace*{-0.5\baselineskip}
\begin{equation}
  \label{eq:3.3}
 \frac{p_\Phi}{c^2} =  \varrho_\Phi- \frac{2V}{c^2} =
  -\frac{(1{+}\gamma)\, \varrho_{\Phi 0}}{3\,x^{2-\gamma}}
  \qquad\mbox{and}\qquad
  w = -\frac{1+\gamma}{3}\,,
\end{equation}
i.e. we obtain a time-independent value of $w$ automatically.
According to Wang et al.~\cite{Wang}, the cosmological constant model
($w{=}{-}1$) has so far still ''the best performance in fitting the
current observational data'' (e.g. for data obtained from Type Ia
supernova observations, see Ref.~\cite{Betoule}), while some of them
``mildly favor'' $w{=}{-}1{-}\epsilon$ with small $\epsilon{>}0$, a
situation which corresponds to a phantom DE with a so called big rip
(see e.g. page 55 of Ref.~\cite{Weinberg1}).

Inserting the ansatz $\gamma{=}2{-}3\epsilon$ with $\epsilon{>}0$ in
Eq.~(\ref{eq:3.3}) we obtain $w{=}{-}1{+}\epsilon$. Since according to
Eqs.~(\ref{eq:1.25}), $\gamma{=}2$ can be approached from below as
closely as wanted, $\epsilon$ can be chosen arbitrarily small. This
means that our model can be fitted to the current observational data
just as well as the cosmological constant model. Choosing,
e.g., $3\,\epsilon{=}0.01$ or $\gamma{=}1.99$ yields the (normalized)
radius $\zeta{=}9.6\cdot 10^{12137}$ of the multiverse and a
corresponding spatial curvature of the order $ 10^{-24276}\!/\!R^2$.
Since $w$ differs from $w{=}{-}1$ only very little at all times, the
space-time generated by the field $\Phi(T)$ approaches period by
period very closely varying de Sitter spaces with different
cosmological constants.

Should forthcoming observations necessitate a value $w\,{<}{-}1$, this
would not completely disqualify the present model, but only its
representation of DE by a scalar field $\Phi$ or the use of
Eqs.~(\ref{eq:0.2})-(\ref{eq:0.3}), respectively. In an appropriately
changed representation the big rip associated with $w\,{<}{-}1$ would
imply an additional constraint, namely that it would have to happen
later than now.

The time behavior of the cosmic acceleration caused by DE is, in
principle, measurable and could turn out as another criterion for the
usefulness of DE models. According to most theoretical models the
acceleration is still increasing. However, in 2009, it was proposed
for the first time, that it ``may have already peaked and that we are
currently witnessing its slowing down.''\cite{Shafieloo} In
Subsection~\ref{sec:infl-sub}, some parameters relevant in this
context were calculated. For the $\gamma$-values close to $2$
following from the above, the quantities $\dddot{y}_0$ and $\dot{q}_0$
specified in Eq~(\ref{eq:2.7}) are both yielding an increase of the
cosmic acceleration. For the areas of the multiverse between the
subuniverses, the situation looks somewhat different. From the
approximation $\dot{x}(\tau){=}\alpha\,x^{\gamma/2}$ with
$\alpha{=}0.826$ and Eq.~(\ref{eq:1.1}b), one easily obtains
\begin{displaymath}
  q = - \frac{A\,\ddot{A}(t)}{\dot{A}^2(t)} = -\frac{\gamma}{2}
  \approx -1
  \qquad\mbox{and}\qquad
  \dot{q}(t) = 0\,.
\end{displaymath}
A local quantity even more closely related to the acceleration
$\ddot{A}(T)$ is the specific acceleration $S{=}\ddot{A}(T)/A$.
For this, we get
\begin{equation}
  \label{eq:3.4}
  S(t) = \frac{\gamma\,\alpha^2\,x^{\gamma-2}}{2t_{H0}^2}
    \qquad\mbox{and}\qquad
    \dot{S}(t) = -\frac{\gamma (1{-}\gamma/2)\alpha^3}{t_{H0}^2\,
      x^{3(1-\gamma/2)}}\,.
\end{equation}
Whereas $\dot{S}(t)$ is negative for the pure DE field of a
$\Phi$-multiverse, the time derivative of corresponding quantity
$s{=}\ddot{a}(t)/a$ in our universe is not much different from
$\dddot{y}_0$ given by Eq.~(\ref{eq:2.7}a) (only $0.441{\to}\,0.723$)
and is positive.

Recently, a comprehensive study on the time behavior of the cosmic
acceleration (CA) was performed, considering both theoretical models
and observational data \cite{Wang2}. It was concluded that ``due to
the low significance, the slowing down of CA is still a theoretical
possibility that cannot be confirmed by the current observations.''
Furthermore, it was found, that ``a flat Universe favors an eternal CA,
while a non-flat Universe prefers a slowing down CA.''  This is, if
only partly, supported by our model:
$\dot{S}_0{=}{-}\gamma(1{-}\gamma/2)\,\alpha^3\!/t_{H0}^2$, the
present value of $\dot{S}(t)$, is zero at $\gamma{=}2$ and assumes its
minimum $\dot{S}_0{=}{-}\alpha^3\!/t_{H0}^2$ at $\gamma{=}1$. Since
according to Eq.~(\ref{eq:1.33}b) or table~\ref{tab:1}, the curvature
$K{=}1/(R\zeta)^2$ of the multiverse increases with decreasing
$\gamma{<}2$, the CA is indeed progressively slowing down with
increasing spatial curvature at least in the interval
$1{\le}\gamma{<}2$.

\newpage

\section{Conclusions}

Special models of a multiverse were analyzed whose main purpose
consists in providing space and time for a multitude of subuniverses
with ours among them. Specific reasons led to assuming closed space
geometry with positive curvature for homogeneous and isotropic
$\Phi$-multiverses which are generated by a scalar quantum field
$\Phi$ and were shown to provide the background and embracing frame of
the gradually emerging subuniverses. Constituting space-times of
similar origin and structure, the $\Phi$-multiverses can be
interpreted as generalizations of de Sitter space, period-wise
approaching varying manifestations of it very closely. The scalar
field $\Phi$, needed for driving an inflation-like expansion of them,
is chosen such that it originates by a creation out of nothing via
quantum-mechanical tunneling. Its energy density $\varrho_\Phi c^2$
can be identified directly with that of the DE in our universe, what
implies that it causes the presently observed acceleration of the
expansion of our universe. A further generalization results from
incorporating primordial matter in the process of creation out of
nothing. Through this, the starting point of spatial expansion is
endowed with the properties of an unstable equilibrium, leading to an
especially efficient initial homogenization of all physical
quantities.  Furthermore the inclusion of primordial matter can be
interpreted such, that together with the ingredients also the
information about their physical properties emerge from the tunneling
process. In this sense of information transmission, the simple matter
density $\varrho_m$ of the multiverse can be understood as an
all-inclusive representative for all kinds of matter to appear later
in subuniverses.

The mass density $\varrho_\Phi$ of the dark energy field $\Phi$ can be
expressed as a monotonic function of the spatial curvature of the
multiverse, decreasing and approaching zero simultaneously with it
(Subsection~\ref{sec:eternal}). This has a significant implication: It
is an important consequence of a famous principle of Mach%
\footnote{Mach's principle can be summarized in short as: \textit{The
    inertia of each mass is caused by its interaction with all other
    masses in the universe}. In developing general relativity,
  Einstein was influenced by this principle. To his disappointment
  (see Ref.~\cite{Pais}), Mach's principle was no general consequence
  of his field equations since, e.g., flat space-time is a
  solution. (Einstein considered de Sitter space in the first place as
  an obstacle [see \textit{Postscript} in Ref.~\cite{de-Sitter1}], and
  for some time, he looked for ways to rule it out.)}%
~(see Ref.~\cite{Mach} or p.~179 and p.~199 in Ref.~\cite{Rebhan1})
that space and time are only meaningful in the presence of matter or
energy. This is automatically satisfied by the current model, because
at all finite times, the spatial curvature and with it $\varrho_\Phi$
is unequal zero. Furthermore, $\varrho_\Phi$ is an indirectly
measurable quantity, and its permanent decay provides an arrow of
time.

It appears worth mentioning that the extension of solutions to
imaginary values of the time and the scalar field $\Phi$ preserves
their property of being exact solutions of the underlying
equations. In the case of $\Phi$, an imaginary value is nothing
unusual because quantum fields can even be complex. It is quite
another matter, that simultaneously, the transition to a regime takes
place, in which a not yet fully available quantum theory of the
gravitational field would be required. Thereby $\Phi$ raises no
problems again, because it is already a quantum quantity. Concerning
the gravitational field represented by $A(T)$, it fits in well that
the corresponding equation, (\ref{eq:1.55}), can be interpreted as the
quasi-classical approximation to a quantum-mechanical equation. In the
case of de Sitter space, this interpretation is supported by
approximate solutions of the Wheeler-De-Witt equation, the
Schr\"odinger equation for stationary wave functions of a universe or
multiverse, in a so-called minisuperspace~\cite{Wheeler, DeWitt,
  Vilenkin2, Linde3}.

Like de Sitter space, the models considered in this paper must be
regarded as toy models primarily chosen because of mathematical
simplicity. They are well-suited for demonstrating the feasibility of
certain physical properties, but nature may well prefer other
models. For this reason Section~\ref{sec:late-ev} started with
employing a more general density
$\varrho_\Phi{=}\varrho_{\Phi 0}f(x)$. It was found in this context
that the initial value $\varrho_i$ of the mass density $\varrho$ is,
independent of its composition and later behavior, already fully
determined by the requirement of zero initial expansion velocity, the
prerequisite for quantum tunneling from nothing. According to
table~\ref{tab:1}, in terms of the age of our universe, the present
age of the underlying space-time is not especially old, a big part of
it being spent for the early evolution ($\tau(x)$ assumes the values
$\tau(1){=}242$, $\tau(10^{-5}){=}228$ and $\tau(10^{-10}){=}216$ for
$\gamma{=}1.99$); in contrast, its present radius assumes extremely
high values so that the spatial curvature lies far below
measurability.

According to Section~\ref{sec:obs-const}, by choosing
$\gamma{=}2{-}3\epsilon$ or $w{=}1{-}\epsilon$ with sufficiently small
$\epsilon{>}0$, our model can be fitted to the current observational
data as well as the cosmological constant model, which has ''the best
performance'' in this respect. In a multiverse, all subuniverses
beyond ours are observationally inaccessible, unless ours has had a
collision with another one~\cite{Ellis}. An observation of this kind
would yield a direct proof of the existence of a multiverse.  In
future, a much more precise observational determination of
$\varrho_{\Phi}(t)$ could potentially yield an indirect proof of the
validity of our model (or modifications of it), the observational
verification of $w{\neq}{-}1$ being a necessary precondition; the same
holds for temporal changes of the cosmic acceleration, because present
observations still admit no unequivocal conclusions.

\appendix

\section{Additional motivation for closed space geometry}
\label{app:1}

In a previous paper~\cite{Rebhan3}, it was shown that in uncurved open
space, the recession of galaxies of our universe, usually interpreted
by an expansion of space, can be explained equivalently by a motion of
the cosmic substrate across radially invariable space, caused by an
explosion-like big bang or by inflation.  The two interpretations are
not in contradiction but are related to each other by a one-to-one
transformation between the specific coordinates to which each of them
is restricted, co-moving FRW coordinates in the case of space
expansion and ``explosion coordinates'' in the case of motion across
radially invariable space.

In a spatially flat universe, the transition from expansion coordinates
$t, r, \vartheta,\varphi$ to explosion coordinates $\tau, \rho,
\vartheta, \varphi$ is accomplished by a transformation $t{=}
t(\rho,\tau)$, $r{=} r(\rho,\tau)$ such that the square of the line
element,
\begin{equation}
  \label{app:1.1}
  ds^2 =  c^2\,dt^2 - a^2(t)\big[dr^2+
  r^2(d\vartheta^2+\sin^2\!\vartheta\,d\varphi^2)\big]
\end{equation}
in FRW coordinates, is transformed into
\begin{equation}
  \label{app:1.2}
  ds^2 =  c^2\,g_{00}(\rho,\tau)\,d\tau^2 - d\rho^2
   +g_\Omega(\rho,\tau)\,(d\vartheta^2+\sin^2\!\vartheta\,d\varphi^2)\,.
\end{equation}
In expansion coordinates, the radial expansion is expressed by the
time-dependence of the length element $dl_r{=}a(t)\,dr$, whereas in
explosion coordinates, there is no radial expansion due to
$dl_\rho{=}d\rho$. Inserting $dt{=}t_\rho d\rho+t_\tau d\tau$ and
$dr{=}r_\rho d\rho+r_\tau d\tau$ in Eq.~(\ref{app:1.1}) and comparing
with Eq.~(\ref{app:1.2}) leads to
\begin{displaymath} 
  c^2t_\rho t_\tau = a^2r_\rho r_\tau\,,\quad c^2t_\rho^2-a^2r_\rho^2=-1\,, 
  \quad  g_{00} = t_\tau^2-a^2r_\tau^2/c^2\,,\quad g_\Omega =-a^2r^2\,.
\end{displaymath}
In Ref.~\cite{Rebhan3}, it was shown that these equations have
solutions observing the conditions $\rho{=}0$ and $t{=}\tau$ at
$r{=}0$.

In a (closed) universe of positive curvature, the square of the line
element in FRW coordinates is%
\vspace*{-0.5\baselineskip}
\begin{equation}
  \label{app:1.3}
   ds^2 = c^2\,dt^2 - a^2(t)\big[d\chi^2+
   \sin^2\chi (d\vartheta^2+\sin^2\vartheta d\varphi^2)\big]\,.
\end{equation}
Since the non-angular part is exactly the same as in a flat universe
(only $r$ being replaced by $\chi$), as far as time and the radial
coordinate are concerned, the transformation to explosion coordinates
can be transferred without alterations. This means that in a closed
universe, the transition to explosion coordinates without radial
expansion is possible as well.  However, a severe restriction must be
made here: Explosion coordinates are meaningful only locally; on a
global scale they would be associated with counter-streams of matter
or galaxies, and, still worse, the radial extent of the universe would
remain unaltered from the very beginning and have to start with
extreme magnitude in order to fit in with present observations.

From this, we infer that in a closed curved multiverse, an intrinsic
global expansion of space exists which cannot be transformed away and
is not present in an uncurved open multiverse. This served as an
additional motivation for employing closed space geometry and endowing
the field $\Phi$ with intrinsic properties.

A different and even simpler approach leads to the same conclusion.
In flat and infinitely extended space, a big bang with or without
inflation is a local event restricted to a region that is initially
extremely small but extends very rapidly. The boundary of the affected
region propagates into its surroundings at the speed of
light. Outside, the situation is the same as before the big bang, and
there is no reason whatever for an expansion of space there. At the
present time $t_0$, the metric radius of the region influenced by the
big bang is $ct_0$ when judged from outside whereas it is appreciably
larger when judged from inside in the system of co-moving FRW
coordinates. (Already the distance of the particle horizon is
$\approx 2.3\,ct_0$.) This means, that the internal and external
coordinate systems cannot be joined in a meaningful manner, and
indicates, that an internal continuation of the external coordinates
must exist which, as the latter ones, exhibit no radial space
expansion.

\section{Slow roll approximation}
\label{app:3}

Slow roll means that the term $\ddot{\Phi}(T)$ on the left hand side
of Eq.~(\ref{eq:0.3}) is dominated by the friction term
${\sim}\dot{\Phi}(T)$, i.e.
$|\ddot{\Phi}(T)/(3H\dot{\Phi}(T))|{\ll}1$. Differentiating the square
of Eq.~(\ref{eq:1.19}) with respect to $T$ and subsequently dividing
it by $2\dot{\Phi}(T)$ yields
\begin{displaymath}
  \ddot{\Phi}(T) = -\frac{\mu c^4 \dot{A}(T)\,(\varrho'_\Phi(A){+}A\varrho''_\Phi(A))}
  {6\hbar^2\dot{\Phi}(T)}
\end{displaymath}
with what we get
\vspace{-0.5\baselineskip}
\begin{displaymath}
  \frac{\ddot{\Phi}(T)}{3H\dot{\Phi}(T)} = \frac{1}{6}\,
  \left(1+\frac{A\varrho''_\Phi(A)}{\varrho'_\Phi(A)}\right)\,.
\end{displaymath}
Inserting in this $\varrho_\Phi(A){=}\varrho_{\Phi 0}A^{\gamma-2}/A_0^{\gamma-2}$ from
Eqs.~(\ref{eq:1.1}a) and~(\ref{eq:1.25}a) yields
\begin{equation}
  \label{app:3.1}
  \left|\frac{\ddot{\Phi(T)}}{3H\dot{\Phi}(T)}\right| = \frac{2{-}\gamma}{6}
  \leq 10^{-2} \qquad\mbox{for}\quad \gamma \geq 1.94\,.
\end{equation}

Neglecting $\ddot{\Phi}(T)$ and using $H{=}\dot{A}(T)/A(T)$,
$V'(\Phi){=}\dot{V}(T)/\dot{\Phi}(T)$ as well as
$\dot{V}(T)/\dot{A}(T){=}V'(A)$, from Eq.~(\ref{eq:0.3}), we obtain
\begin{displaymath}
  \frac{3\dot{A}(T)\,\dot{\Phi}(T)}{A(T)} = -\frac{\mu c^2}{\hbar^2}\,V'(\Phi)
  = -\frac{\mu c^2\,\dot{V}(T)}{\hbar^2\,\dot{\Phi}(T)}
  \qquad\mbox{or}\qquad
  \dot{\Phi}^2(T) = -\frac{\mu c^2\,A\,V'(A)}{3\,\hbar^2}\,. 
\end{displaymath}
With Eq.~(\ref{eq:1.19}), the last equation becomes
$V'(A){=}c^2\,\varrho'_\Phi(A)$. Choosing an
integration constant such that $V{\to}\,0$ for $A{\to}\,\infty$, by
integration and with $A{=}A_0x$, we finally obtain%
 \begin{equation}
   \label{app:3.2}
   V(x) = c^2\,\varrho_\Phi(x) = \varrho_{\Phi 0}\,c^2\,x^{\gamma-2}\,.
 \end{equation}

\end{document}